\documentclass[journal]{IEEEtran}
\ifCLASSINFOpdf
\else
\fi
\usepackage{amsmath,epsfig, psfrag, amstext,amsfonts,amssymb,subcaption,color,bm,amsthm,algorithm,algorithmic, multirow, tabularx}
\usepackage[outdir=./]{epstopdf}
\usepackage{cite}
\usepackage[capitalize]{cleveref}

\usepackage[showonlyrefs]{mathtools}

\usepackage{tikz, pgfplots,pgfplotstable}
\usetikzlibrary{calc,positioning,plotmarks,shapes,snakes}
\pgfplotsset{compat=newest}

\pgfplotsset{select coords between index/.style 2 args={
    x filter/.code={
        \ifnum\coordindex<#1\fi
        \ifnum\coordindex>#2\fi
    }
}}

\usepackage{pifont}

\newtheorem*{remark}{Remark}

\newcolumntype{L}[1]{>{\raggedright\arraybackslash}p{#1}}
\newcolumntype{C}[1]{>{\centering\arraybackslash}p{#1}}
\newcolumntype{R}[1]{>{\raggedleft\arraybackslash}p{#1}}

\usepackage[utf8]{inputenc}

\newcommand{\myset}{\ensuremath{S}}

\begin{document}
\title{Bayesian Cooperative Localization Using Received Signal Strength With Unknown Path Loss Exponent:\\ Message Passing Approaches}

\author{Di~Jin,
        Feng~Yin,
        Carsten~Fritsche,
        Fredrik~Gustafsson,
        and~Abdelhak~M.~Zoubir,% <-this % stops a space
\thanks{D.~Jin and A.~M.~Zoubir are with the Signal Processing Group at Technische Universität Darmstadt, Darmstadt, Germany (correspondence e-mail: djin@spg.tu-darmstadt.de).}% <-this % stops a space
\thanks{F.~Yin is with the School of Science and Engineering at Chinese University of Hong Kong, Shenzhen, China.}% <-this % stops a space
\thanks{C.~Fritsche and F.~Gustafsson are with the Department of Electrical Engineering at Linköping University, Linköping, Sweden.}% <-this % stops a space
% \thanks{Manuscript received April 19, 2005; revised August 26, 2015.}
}


\markboth{submitted to XXXXXX for publication}%
{}
% {Shell \MakeLowercase{\textit{et al.}}: Bare Demo of IEEEtran.cls for IEEE Journals}



\maketitle

\begin{abstract}
We propose a Bayesian framework for the received-signal-strength-based cooperative localization problem with unknown path loss exponent. 
Our purpose is to infer the marginal posterior of each unknown parameter: the position or the path loss exponent. This probabilistic inference problem is solved using message passing algorithms that update messages and beliefs iteratively.
To enable the numerical tractability, we combine the variable discretization and Monte-Carlo-based numerical approximation schemes. 
To further improve computational efficiency, we develop an auxiliary importance sampler that updates the beliefs with the help of an auxiliary variable.
To sample from a normalized likelihood function, which is an important ingredient of the proposed auxiliary importance sampler, we develop a stochastic sampling strategy that mathematically interprets and corrects an existing heuristic strategy.
The proposed message passing algorithms are analyzed systematically in terms of computational complexity, demonstrating the computational efficiency of the proposed auxiliary importance sampler. 
Various simulations are conducted to validate the overall good performance of the proposed algorithms.
\end{abstract}

\begin{IEEEkeywords}
Belief propagation, cooperative localization, message passing, received signal strength, stochastic sampling.
\end{IEEEkeywords}

% \fi
\IEEEpeerreviewmaketitle

% \section{Introduction}
% for IEEE journal papers produced under \LaTeX\ using
% IEEEtran.cls version 1.8b and later.
\section{Introduction}
\IEEEPARstart{R}{ecently}, wireless cooperative localization has attracted much interest. In cooperative localization \cite{Patwari2006,Wymeersch2009,LI2018,Wymeersch2018}, all internode measurements can be exploited, leading to many appealing advantages, among others, expanding the capabilities of locating positions without ambiguity and improving the performance on estimation accuracy. The benefits of cooperation among nodes have been theoretically demonstrated in \cite{Shen2010,Buehrer2016}.
Depending on whether the localization problem is formulated in a probabilistic manner, the existing algorithms for cooperative localization can be categorized into deterministic and probabilistic approaches. In the first category, the positions (and model parameters if any) are assumed to be deterministic but unknown, and only a deterministic point estimate is provided for each unknown parameter. Classical approaches, to mention some, include the maximum likelihood (ML) approach \cite{Patwari2006}, convex-optimization-based algorithms \cite{Tseng2007,Ouyang2010,Wang2011,Vaghefi2013,Simonetto2014,Tomic2015}, multidimensional scaling (MDS) \cite{Costa2006,Li2007} and expectation-conditional maximization (ECM) \cite{Yin2015}.  
On the other hand, the class of Bayesian approaches treat the positions as random variables and formulate cooperative localization as a probabilistic inference problem. These approaches take advantage of prior information of parameters. Most importantly, the posterior distribution of each position is inferred, which contains much more information than just one deterministic point estimate, e.g., the modality of the position and its associated uncertainty. Representative probabilistic approaches include the nonparametric belief propagation (NBP) \cite{Sudderth2003,Ihler2005}, sum-product algorithm over a network (SPAWN) \cite{Wymeersch2009} and their low-complexity variants \cite{Lien2012,Li2015,Jin2015,Scheidt2016}. 

%% why RSS
Among different position-related signal metrics, received signal strength (RSS) has gained much attention due to its ubiquitousness in wireless radio frequency signals \cite{Viani2011}. For instance, an RSS indicator (RSSI) has been encoded in the IEEE 802.15.4 standards \cite{IEEE_802_15}. Despite its comparatively high uncertainty about position, RSS measurement can be exploited to enable low-cost, simple and opportunistic localization systems, without the need of additional hardware. However, many existing works on RSS-based localization, such as \cite{Ouyang2010,Patwari2003}, are based on the assumption that the classical path loss propagation model is perfectly known. This oversimplified assumption is impractical for two reasons. Firstly, the estimation of these model parameters usually relies on a laborious calibration phase, where a large amount of training data needs to be collected and processed. Such a calibration step is, however, very time consuming and even impossible in many scenarios, such as monitoring and surveillance applications in hostile or inaccessible environments \cite{Li2006}. Secondly, these model parameters, particularly the path loss exponent (PLE), are time varying, due to the changing environment, e.g., weather conditions or human behaviors \cite{Mao2007,Rappaport2001}. Without a frequent recalibration, the resulting mismatch will significantly deteriorate the localization performance. In order to overcome this problem, these model parameters should be assumed unknown and jointly estimated with the positions. 

In this paper, we focus on the case with unknown PLE for the reason that a slight deviation of PLE may severely deteriorate the localization performance, as theoretically and algorithmically demonstrated in \cite{Salman2012a,Salman2012b}. For the case of noncooperative localization, there exist several works dealing with unknown PLE. In \cite{Li2006}, the target position and the PLE are estimated jointly by solving an ML problem using the Levenberg-Marquardt algorithm. In \cite{Salman2012b, Chan2011}, the ML problem is first relaxed by linearizing the problem and then simplified by replacing the position variable with a function of the PLE variable. By doing so, the cost function depends only on the one-dimensional ($1$D) PLE variable, and the resulting optimization problem can be readily solved using grid search. 
In \cite{Li2006,Gustafsson2012}, the location is estimated by eliminating the nuisance parameter: the PLE parameter (or several other model parameters).
The original ML problem in \cite{Li2006} is simplified by representing the PLE as a function of the position variable in \cite{Salman2012a}. 
In \cite{Yin2013}, along with several model parameters, the location is estimated based on the expectation and maximization criterion.
% and in \cite{Gustafsson2012} the location is estimated by treating the model parameters as nuisance parameter, which are eliminated using the principle of separable least squares. 
In \cite{So2011,Wang2012}, the location and the PLE are estimated in an alternating manner. More precisely, the position is estimated based on an initialized (or estimated) PLE, and afterwards the PLE is estimated based on the updated position estimate. This procedure iterates until certain termination condition is met. In the cooperative case, RSS-based localization with an unknown PLE is even more challenging. To the best of our knowledge, only very limited works exist, including \cite{Wang2012,Vaghefi2013,Tomic2015}, where the alternating strategy is adopted to handle the unknown PLE, like in the noncooperative case. In our view, despite its straightforwardness and simplicity, such an alternating strategy is quite heuristic and lack of theoretical support. 

%%%%%%%%%%%%%%%%%%%%%%%%%%%%%%%%%%%%%%%%%%%%%%%%%%%%%%%%%%%%%%%%%%%%%%%%%%%%%%%%%%%%%%%%%%%%%%%%%%%%%%%%%%%%%%%%%%%%%%%%%%%%%%%%%%%%%%%%%%%%%%%%%%%%%%
Different from the existing works, we treat the PLE as a random variable and formulate the problem in a Bayesian framework. The reasons are as follows. First, when the PLEs between different propagation links differ, a random variable characterizing the averaging behavior of the collection of all PLEs is more suitable than just one deterministic PLE value. Second, characterizing the PLE as a random variable enables us to integrate any prior information, if available, into the parameter estimation. 
Under the Bayesian umbrella, the cooperative localization problem with unknown PLE becomes a probabilistic inference problem. In this problem, we derive message passing algorithms to infer the marginalized posterior distribution of each unknown parameter: the position or the PLE. 
To enable mathematical tractability, we combine the variable discretization and Monte-Carlo-based numerical approximation mechanisms. 
In addition, to reduce the computational complexity, we propose an auxiliary importance sampler for belief update that has a complexity order scaling linearly with the number of samples.
Moreover, we develop a novel strategy for sampling from a normalized likelihood function, which plays an important role in the auxiliary importance sampler and mathematically interprets and corrects an existing heuristic sampling strategy. 
The proposed sampling strategy will benefit many existing works, such as \cite{Ihler2005,Wymeersch2009}, since this task is an embedded step in many message-passing-based cooperative localization algorithms.

This paper is organized as follows: In \cref{sec:problem formulation}, we formulate the RSS-based cooperative localization problem with unknown PLE mathematically. Fundamental concepts in message passing algorithms are given in \cref{sec:fundamentals}. We discuss how to approximate the messages in \cref{sec:message update} and demonstrate how to update the beliefs approximately in \cref{sec:message multiplication}. Some important issues are discussed in \cref{sec:other issues}. The proposed algorithms are evaluated using extensive simulations in \cref{sec:simulation results}. Finally, \cref{sec:conclusion} concludes the paper.

\emph{Notation}: Throughout this paper, boldface lowercase letter $\mathbf{x}$ is reserved for vector. $\lVert \cdot \rVert$ stands for the Euclidean norm, and $| \cdot |$ denotes the cardinality of a set. $\mathcal{N}(\mu, \sigma^2)$ denotes a Gaussian distribution with mean $\mu$ and variance $\sigma^2$; $\mathcal{U}\left[a, b\right)$ denotes a uniform distribution with two boundaries $a$ and $b$; $\text{log}\mathcal{N}(\mu, \sigma^2)$ denotes a log-normal-distributed random variable $x$ with $\mu$ and $\sigma^2$ being the mean and variance of $\text{log}x$. $f(\cdot)$ and $p(\cdot)$ are reserved for the probability density function (pdf) and the probability mass function, respectively, $f_{\mathcal{N}}$ for the pdf of a Gaussian distribution. The notation $ \Gamma \backslash i $ represents a set consisting all elements in the set $\Gamma$ excluding the element $i$. $\{ x^l \}_{l = 1}^L$ is a short notation for a collection of samples $\left\lbrace x^1, \ldots, x^L\right\rbrace$.
%%%%%%%%%%%%%%%%%%%%%%%%%%%%%%%%%%%%%%%%%%%%%%%%%%%%%%%%%%%%
\section{Problem Formulation} 
\label{sec:problem formulation}
Consider a wireless sensor network (WSN) in 2D space with two types of nodes: blindfolded nodes with unknown locations and reference nodes with known locations, referred to as \emph{agents} and \emph{anchors}, respectively. Let $\mathbf{x}_{i}=[x_i, y_i]^{T}$ denote the location of each node, where $i \in \myset_u = \left\lbrace 1, \ldots, N_u\right\rbrace$ as to an agent and $i \in \myset_a = \left\lbrace N_u+1, \ldots, N\right\rbrace$ as to an anchor. The index set of all nodes is denoted by $S$, and we have $S = S_u \bigcup S_a$. If there exists communication between two sensor nodes $i$ and $j$, then they are neighbors. We denote the index set of node $i$'s neighbors by $\Gamma_i$. 

Using the well known log-distance path loss propagation model, the RSS measurement $r_{ij}$, coming from node $i$ and received by node $j$, is given by 
\begin{equation}
r_{ij} = A_{i} - 10 \alpha \text{log}_{10}(d_{ij}/d_0) + v_{ij}, 
\label{eq:measurement model}
\end{equation}      
where $d_0$ is a predefined reference distance; $A_{i}$ denotes the reference power in dBm at $d_0$, and it is assumed to be known; 
%which can be realized by reporting its transmit power, 
$\alpha$ denotes the path loss exponent (PLE) that is assumed unknown; $d_{ij} \triangleq \lVert \mathbf{x}_{i} - \mathbf{x}_{j} \rVert $ is the 
Euclidean distance; $v_{ij}$ stands for the log-normal shadowing error that is modeled by $v_{ij} \sim \mathcal{N}(0, \sigma_{ij}^2)$. A symmetric propagation is considered, meaning that we make no difference between the measurements $r_{ij}$ and $r_{ji}$. 
The collection of all RSS measurements is denoted by $\mathbf{r} \triangleq \left\lbrace r_{ij}: (i,j)\in\Gamma \right\rbrace$, where $(i,j)$ represents that nodes $i$ and $j$ are neighbors, and $\Gamma \triangleq \left\lbrace (i, j) : j\in\Gamma_i \text{ and } j>i; i\in \myset_u \right\rbrace$ denotes the set of all pairs of neighboring nodes. In alignment with the majority of the existing works, we assume that these shadowing measurement errors $v_{ij}$ for all $(i,j) \in \Gamma$ are independent. The distribution of $v_{ij}$, denoted by $f_{v_{ij}}(v_{ij})$, is assumed to be known. 

From a Bayesian perspective, we treat the PLE $\alpha$ and each position $ \mathbf{x}_i , i\in \myset $, as random variables, whose prior distributions are denoted by $f(\alpha)$ and $f(\mathbf{x}_i), i\in \myset$, respectively. All positions and the PLE variable are assumed to be mutually independent, i.e., $f(\alpha, \mathbf{x}_1, \ldots, \mathbf{x}_{N}) = f(\alpha)\cdot f(\mathbf{x}_1)\cdots f(\mathbf{x}_{N})$. Our purpose is to infer the marginalized posterior distribution (marginal posterior) of each unknown parameter, which is $f(\alpha | \mathbf{r})$ or $f(\mathbf{x}_i | \mathbf{r})$, $i \in \myset_{u}$, from the measurements $\mathbf{r}$ and the prior information about all parameters. 

%%%%%%%%%%%%%%%%%%%%%%%%%%%%%%%%%%%%%%%%%%%%%%%%%%%%%%%%%%%%%%%%%%%%%%%%%%%%%%%%%%%%%%%%%%%%%%%%%%%%%%%%%%%
\section{Fundamentals on Cooperative Localization via Message Passing}
\label{sec:fundamentals}
To infer the marginal posterior of the PLE variable $\alpha$ and that of each position $\mathbf{x}_i, i\in \myset_u$, 
we start with the joint posterior distribution $f(\mathbf{x}_1, \ldots, \mathbf{x}_N, \alpha \left|\mathbf{r} \right.)$. Under the assumptions made in the preceding section, it has the form of
\begin{align*}
 f(\mathbf{x}_1, \ldots, \mathbf{x}_N, \alpha \left|\mathbf{r} \right.) \hspace{-1mm} \propto f(\alpha)\prod_{i = 1}^{N} \left( \hspace{-1mm} f(\mathbf{x}_i) \hspace{-3mm}  \prod_{j\in \Gamma_{i}, j>i } \hspace{-3mm} f(r_{ij} \left| \mathbf{x}_i, \mathbf{x}_j, \alpha \right.)\right) \hspace{-1mm} .
\end{align*}
Intuitively, the marginal posterior, say $f(\mathbf{x}_i \left| \mathbf{r} \right.)$, can be calculated as follows: 
\begin{align*}
f(\mathbf{x}_i \left| \mathbf{r} \right.) = \int \cdots \int f(\mathbf{x}_1, \ldots, \mathbf{x}_N, \alpha \left|\mathbf{r} \right.) ~\mathrm{d}\mathbf{x}_{1:N \backslash i} \mathrm{d}\alpha.
\end{align*}
However, this is intractable due to the high dimensionality of the problem. A well-known local message passing algorithm, called belief propagation (BP), enables the marginalization in an elegant fashion \cite{Pearl88}. 
In the BP, a set of messages are calculated in an iterative manner, and each marginal posterior can be calculated (or approximated) based on a certain set of messages. More details on the BP can be found in \cite{Pearl88}. Despite the fact that several works, e.g., \cite{Sudderth2003, Ihler2009}, exist for cooperative localization via BP, they do not directly apply to our problem. The reason is that unlike a pairwise potential function in the existing works, here the likelihood function $f(r_{ij} \left| \mathbf{x}_i, \mathbf{x}_j, \alpha \right.)$ in our problem is of order three. This makes the BP algorithm for our problem not straightforward, and, hence, we will derive it explicitly in what follows. 
We first represent the joint posterior distribution $ f(\mathbf{x}_1, \ldots, \mathbf{x}_N, \alpha \left|\mathbf{r} \right.) $ using a factor graph, see \cref{fig:belief propagation}. There are two distinctive nodes in the factor graph, the variables in circles and the factors in squares, representing the random variables and the likelihood functions (or prior distributions), respectively. Two position variables are connected via a factor if there is a measurement between them available. The PLE variable is connected to all likelihood functions as it is related to all measurements.
 
The key idea of the BP is to update a set of messages iteratively, which contribute to calculating the marginal posteriors. Using $f_{ij}$ as a short-hand notation for the likelihood function $f(r_{ij} \left| \mathbf{x}_i, \mathbf{x}_j, \alpha \right.)$, we denote the message from factor $f_{ij}$ to variable $\alpha$ by $m_{f_{ij} \rightarrow \alpha}(\alpha)$ and that from $f_{ij}$ to $\mathbf{x}_i$ by $m_{f_{ij} \rightarrow \mathbf{x}_i}(\mathbf{x}_i)$. The messages $m_{f_{ij} \rightarrow \alpha}(\alpha)$ and $m_{f_{ij} \rightarrow \mathbf{x}_i}(\mathbf{x}_i)$ are updated according to the following rule:
\begin{subequations}
\label{eq:message integration original}
\begin{align}
 m_{f_{ij} \rightarrow \alpha}^{n}(\alpha) & \propto \iint f(r_{ij} \left|\mathbf{x}_i, \mathbf{x}_j, \alpha \right.) \; f(\mathbf{x}_i) \hspace{-1mm} \prod_{s \in \Gamma_i \backslash j} \hspace{-1mm} m_{f_{si} \rightarrow \mathbf{x}_i}^{n-1}(\mathbf{x}_i) \nonumber \\
                                        & \cdot f(\mathbf{x}_j) \prod_{t \in \Gamma_j \backslash i} m_{f_{tj} \rightarrow \mathbf{x}_j}^{n-1}(\mathbf{x}_j) ~\mathrm{d}\mathbf{x}_i  ~\mathrm{d}\mathbf{x}_j, \label{eq:message integration original 2} \\
 \hspace{-2 em} m_{f_{ij} \rightarrow \mathbf{x}_i}^{n}(\mathbf{x}_i) 
                                                    & \propto \iint f(r_{ij} \left|\mathbf{x}_i, \mathbf{x}_j, \alpha \right.) \; f(\mathbf{x}_j)  \hspace{-1mm} \prod_{t \in \Gamma_j \backslash i}  \hspace{-1mm} m_{f_{tj} \ \rightarrow \mathbf{x}_j}^{n-1}(\mathbf{x}_j) \nonumber \hfill \\
                                                    &\cdot f(\alpha) \hspace{-4mm} \prod_{(u,z) \in \Gamma\backslash (i,j)} \hspace{-4mm} m_{f_{uz} \rightarrow \alpha }^{n-1}(\alpha) ~\mathrm{d}\mathbf{x}_j ~ \mathrm{d}\alpha . \label{eq:message integration original 1} 
\end{align}
\end{subequations}
Here, the superscript $n$ is the iteration index, $\Gamma_i \backslash j$ denotes the set of all neighbors of node $i$ excluding node $j$, and $\Gamma\backslash (i,j)$ denotes the set of all pairs of neighboring nodes excluding the pair $(i, j)$. To facilitate compact notation, we will simplify $m_{f_{ij}\rightarrow \alpha}(\alpha)$ and $m_{f_{ij}\rightarrow \mathbf{x}_i}(\mathbf{x}_i)$ to $m_{ij}(\alpha)$ and $m_{ij}(\mathbf{x}_i)$, respectively. An illustrative explanation of \cref{eq:message integration original 1} is depicted in Fig. \ref{fig:belief propagation}, where the messages enclosed in the dashed circle contribute to calculating the message $m_{ij}(\mathbf{x}_i)$. At the first glance, the message update rule in \cref{eq:message integration original} seems tedious. In the subsequent context, a reformulation of \cref{eq:message integration original} will be given in \cref{eq:message integration}, therewith facilitating the interpretation of the messages $m_{ij}(\alpha)$ and $m_{ij}(\mathbf{x}_i)$. 
Based on these messages, the marginal posteriors (referred to as \emph{beliefs}) can be, either exactly or approximately, calculated. More precisely, in each iteration, the beliefs are updated by performing 
\begin{subequations}
 \label{eq:message multiplication}
\begin{align}
 B^{n}(\alpha)         & \propto f(\alpha) \prod_{\left( i,j\right) \in \Gamma } m_{ij}^{n}(\alpha),  \label{eq:message multiplication 1}\\
 B^{n}(\mathbf{x}_i) & \propto f(\mathbf{x}_i) \prod_{j\in\Gamma_{i}} m_{ij}^{n}(\mathbf{x}_i). \label{eq:message multiplication 2}
 \end{align}
\end{subequations}
Here, $B^n(\alpha)$ and $B^n(\mathbf{x}_i)$  denote the belief of the PLE variable $\alpha$ and the belief of the position variable $\mathbf{x}_i$ in the $n$-th iteration, respectively. The belief update rule in \cref{eq:message multiplication}, say $ B^{n}(\mathbf{x}_i)$, can be interpreted as multiplying the messages coming from all factors connected to $\mathbf{x}_i$. As an illustrative example, the belief update rule for $B(\mathbf{x}_i)$ is depicted in Fig. \ref{fig:belief propagation}, where the messages enclosed in the dotted circle contribute to updating $B(\mathbf{x}_i)$. 
% 
%%%%%%%%%%%%%%%%%%%%%%%%%%%%%%%%%%%%%%%%%%%%%%%%%%%%%%%%%%%%%%%%%%%%%%%%%%%%%%%%
%  Define node, factor styles and arrows parallel to lines 
\tikzstyle{FGnode} = [circle, draw, text centered]
\tikzstyle{FGfactor} = [rectangle, draw, text centered]
\tikzset{
parallel segment/.style={
   segment distance/.store in=\segDistance,
   segment pos/.store in=\segPos,
   segment length/.store in=\segLength,
   to path={
   ($(\tikztostart)!\segPos!(\tikztotarget)!\segLength/2!(\tikztostart)!\segDistance!90:(\tikztotarget)$) -- 
   ($(\tikztostart)!\segPos!(\tikztotarget)!\segLength/2!(\tikztotarget)!\segDistance!-90:(\tikztostart)$)  \tikztonodes
   }, 
   % Default values
   segment pos=.5,
   segment length=5ex,
   segment distance=1.5mm,
},
}

%% belief propagation
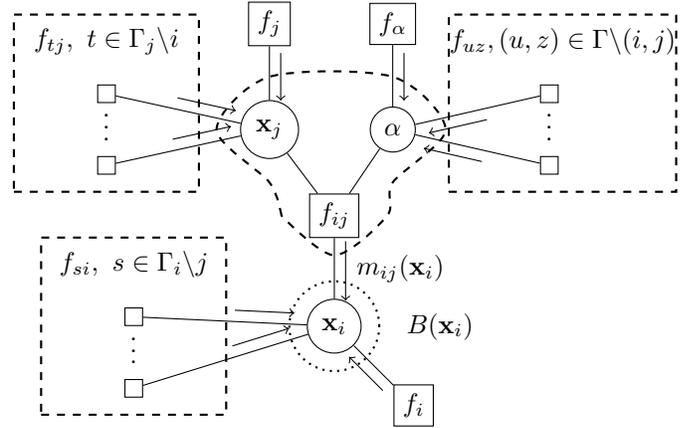
\begin{figure}[t]
 \centering
 \begin{tikzpicture}
  \matrix(FG) [row sep = 1.3em, column sep = 0.3 em]
  {
  \node[FGnode] (xj) {$\mathbf{x}_j$}; & & \node[FGnode] (alpha) {$\alpha$}; \\
  &     \node[FGfactor](gij){$f_{ij}$}; & \\ [1em]
  & \node[FGnode] (xi) {$\mathbf{x}_i$}; & \\
  };
    
  \matrix (neighbor of xj) [row sep = 2em, left = of xj, xshift = -1.5em] 
  {
  \node[FGfactor] (neighbor1 of xj) {}; \\
  \node[FGfactor] (neighbor3 of xj) {}; \\
  };
  
  \matrix (neighbor of alpha) [row sep = 2em, right = of alpha, xshift = 1.5em]
  {
  \node[FGfactor] (neighbor1 of alpha) {}; \\
  \node[FGfactor] (neighbor3 of alpha) {}; \\
  };
  
  \matrix (neighbor of xi) [row sep = 2em, left = of xi, xshift = -3em, yshift = -1em]    % row sep = 4em, yshift = -3em
  {
  \node[FGfactor] (neighbor1 of xi) {}; \\
  \node[FGfactor] (neighbor3 of xi) {}; \\
  };
  
%  priors of all nodes
\node[FGfactor, xshift = 3em, yshift = -3em] (prior of xi) at (xi) {$f_i$};
\node[FGfactor, xshift = 0em, yshift = 4em] (prior of xj) at (xj) {$f_j$};
\node[FGfactor, xshift = 0em, yshift = 4em] (prior of alpha) at (alpha) {$f_{\alpha}$};
  
%   connecting nodes and factor
  \draw (xj) -- (gij);
  \draw (xi) -- (gij);
  \draw (alpha) -- (gij);
  
  \draw (xi) -- (prior of xi);
  \draw (xj) -- (prior of xj);
  \draw (alpha) -- (prior of alpha);

%   arrows above and below links
  \draw [->, black] (gij) to[parallel segment, segment pos = .5] (xi);
  \draw [->] (prior of xj) to [parallel segment, segment pos = 0.5, segment length=4ex ] (xj);
  \draw [->] (prior of xi) to [parallel segment, segment pos = 0.5, segment length=4ex] (xi);
  \draw [->] (prior of alpha) to [parallel segment, segment pos = 0.5, segment length=4ex] (alpha);

%   generate the ellipse symbol for xi's neighbors
  \path (neighbor1 of xi) -- (neighbor3 of xi) node [midway, sloped] {$\dots$};
  \path (neighbor1 of xj) -- (neighbor3 of xj) node [midway, sloped] {$\dots$};
  \path (neighbor1 of alpha) -- (neighbor3 of alpha) node [midway, sloped] {$\dots$};
  
%   generate the rectangle box around xi's neighbors
  \draw[thick, dashed] ($(neighbor1 of xi) + (-3.5em, 3em)$) rectangle ($(neighbor3 of xi) + (3.5em, -1em)$);
  \node at ($(neighbor1 of xi)+(0, 2em)$) {$f_{si}, ~  s \in \Gamma_i \backslash j$};
  \draw[thick, dashed] ($(neighbor1 of xj) + (-3.5em, 3em)$) rectangle ($(neighbor3 of xj) + (3.5em, -1em)$);
  \node at ($(neighbor1 of xj)+(0, 2em)$) {$f_{tj}, ~  t \in \Gamma_j \backslash i$};
  \draw[thick, dashed] ($(neighbor1 of alpha) + (-3.8em, 3em)$) rectangle ($(neighbor3 of alpha) + (4.8em, -1em)$);
  \node at ($(neighbor1 of alpha)+(0.5em, 2em)$) {$f_{uz}, (u,z)\in \Gamma\backslash (i,j)$};

%   generate the connection between xi and its neighbors
  \draw (neighbor1 of xi) -- (xi);
  \draw (neighbor3 of xi) -- (xi);
  \draw [->] (neighbor1 of xi) to[parallel segment, segment pos = .65] (xi);
  \draw [->] (neighbor3 of xi) to[parallel segment, segment pos = .65] (xi);  
  
%   generate the connection between xj and its neighbors
  \draw (neighbor1 of xj) -- (xj);
  \draw (neighbor3 of xj) -- (xj);
  \draw [->] (neighbor1 of xj) to[parallel segment, segment pos = .6] (xj);
  \draw [->] (neighbor3 of xj) to[parallel segment, segment pos = .6] (xj);
  
%   generate the connection between xj and its neighbors
  \draw (neighbor1 of alpha) -- (alpha);
  \draw (neighbor3 of alpha) -- (alpha);
  \draw [->] (neighbor1 of alpha) to[parallel segment, segment pos = .6] (alpha);
  \draw [->] (neighbor3 of alpha) to[parallel segment, segment pos = .6] (alpha);
      
%   placing the mathematical symbols
 \node (message) at (gij) [xshift = 2.5em, yshift = -2em] {\textcolor{black}{$m_{ij}(\mathbf{x}_i)$}};
 \node (belief) at (xi) [xshift = 4em, yshift = 0em] {\textcolor{black}{$B(\mathbf{x}_i)$}};
% \node(xi to gij) at (gij) [xshift = -3.5em, yshift = -3em] {$ m_{\mathbf{x}_i \rightarrow f_{ij}}(\mathbf{x}_i) $};
% \node(xj to gij) at (gij) [xshift = -5em, yshift = 1.5em] {$ m_{\mathbf{x}_j \rightarrow f_{ij}}(\mathbf{x}_j) $};
% \node(alpha to gij) at (gij) [xshift = 5.5 em, yshift = 1.5em] {$ m_{\alpha \rightarrow f_{ij}}(\alpha) $};

%  draw circle for the belief 
\draw [thick, dotted] (xi) circle (1.7em);

%  draw closed shape for the message
% \draw [thick, dashed] ($(gij) + (-2em, 0em)$) -- ( $(xj) + (0em , -2em) $) -- ();
\draw [dashed, thick] plot [smooth, tension=0.5] coordinates { ($(gij) + (-2em, 0em)$) ( $(xj) + (0em , -2em) $) ( $(xj) + (-2em , 0em) $) ( $(xj) + (0em , 2em) $) ( $(alpha) + (0em , 2em) $) ( $(alpha) + (2em , 0em) $) ( $(alpha) + (0em , -2em) $) ( $(gij) + (2em , 0em) $) ( $(gij) + (0em , -1.5em) $) ($(gij) + (-2em, 0em)$)};   
   
 \end{tikzpicture}
\caption{\small{An illustration of factor graph and belief propagation. For clarity, the overlap between three blocks are omitted. Here, $f_i$ and $f_{\alpha}$ are short notations for $f(\mathbf{x}_i)$ and $f(\alpha)$, respectively and $f_{ij}$ for the likelihood function $f(r_{ij} \left| \mathbf{x}_i, \mathbf{x}_j, \alpha \right.)$.}}
\label{fig:belief propagation}
\end{figure}
% 
%%%%%%%%%%%%%%%%%%%%%%%%%%%%%%%%%%%%%%%%%%%%%%%%%%%%%%%%%%%%%%%%%%%%%%%%%%%%%%%% 
% 
Comparing \cref{eq:message integration original} with \cref{eq:message multiplication}, it is obvious that certain terms in \cref{eq:message integration original} can be replaced by \cref{eq:message multiplication}. By doing so, the message update rule in \cref{eq:message integration original} can be equivalently rewritten into a simpler form, namely,
\begin{subequations}
\label{eq:message integration}
 \begin{align}
 \hspace{-0em} m_{ij}^{n}(\alpha)             & \propto \iint f(r_{ij} \left|\mathbf{x}_i, \mathbf{x}_j, \alpha \right.) \frac{B^{n-1}(\mathbf{x}_i)}{m_{ij}^{n-1}(\mathbf{x}_i)}  \frac{B^{n-1}(\mathbf{x}_j)}{m_{ij}^{n-1}(\mathbf{x}_j)} \mathrm{d}\mathbf{x}_i  \mathrm{d}\mathbf{x}_j, \label{eq:message integration 2} \\
 \hspace{-0em} m_{ij}^{n}(\mathbf{x}_i)  & \propto \iint f(r_{ij} \left|\mathbf{x}_i, \mathbf{x}_j, \alpha \right.) \frac{B^{n-1}(\mathbf{x}_j)}{m_{ij}^{n-1}(\mathbf{x}_j)}  \frac{B^{n-1}(\alpha)}{m_{ij}^{n-1}(\alpha)} ~\mathrm{d}\mathbf{x}_j ~ \mathrm{d}\alpha . \label{eq:message integration 1}
 \end{align}
\end{subequations}
Such a reformulation results in a succinct message update rule, and the underlying meaning of the messages becomes better revealed in \cref{eq:message integration}. Taking $m_{ij}^{n}(\mathbf{x}_i)$ as an example, it implies that certain information on $\mathbf{x}_i$ can be inferred from the likelihood function $f(r_{ij} \left|\mathbf{x}_i, \mathbf{x}_j, \alpha \right.)$, given the beliefs of $\mathbf{x}_j$ and $\alpha$. Roughly speaking, the message $m_{ij}(\mathbf{x}_i)$ can be deemed as the information on $\mathbf{x}_i$ coming from its neighbor $j$. 
Alternatively, following the idea in \cite{Wymeersch2009}, the messages can be approximated by ignoring the denominator terms in \cref{eq:message integration}, giving rise to the following message update rule: 
\begin{subequations}
\label{eq:SPAWN_message update}
\begin{align}
 m_{ij}^{n}(\alpha)              & \propto \iint f(r_{ij} \left|\mathbf{x}_i, \mathbf{x}_j, \alpha \right.) B^{n-1}(\mathbf{x}_i) B^{n-1}(\mathbf{x}_j) ~\mathrm{d}\mathbf{x}_i  ~\mathrm{d}\mathbf{x}_j , \label{eq:SPAWN_message integration 2}\\
 m_{ij}^{n}(\mathbf{x}_i)  & \propto \iint f(r_{ij} \left|\mathbf{x}_i, \mathbf{x}_j, \alpha \right.) B^{n-1}(\mathbf{x}_j) B^{n-1}(\alpha) ~\mathrm{d}\mathbf{x}_j ~ \mathrm{d}\alpha . \label{eq:SPAWN_message integration 1}
 \end{align}
\end{subequations}
In this paper, the message passing algorithm in light of \cref{eq:message multiplication,eq:message integration} is referred to as the BP; while that in light of \cref{eq:message multiplication,eq:SPAWN_message update} is referred to as the SPAWN. Note that the difference between the BP and the SPAWN lies in the message update rule. As will be shown later in \cref{sec:computational complexity}, the SPAWN message update rule according to \cref{eq:SPAWN_message update} achieves a significant reduction in computational complexity.

Clearly, the gist of the message passing algorithms is to perform two steps iteratively: updating the messages according to \cref{eq:message integration} (or \cref{eq:SPAWN_message update}) and updating the beliefs according to \cref{eq:message multiplication}. 
To give an overview, we summarize the resulting framework for inferring the marginal posteriors $f(\alpha \left| \mathbf{r} \right.)$ and $f(\mathbf{x}_i \left| \mathbf{r}\right.), \; i \in \myset_u$,
in \cref{alg:centralized}. 
First, we initialize the beliefs, $B^0(\alpha)$ and $B^0(\mathbf{x}_i), i \in \myset$. Here, one sensible choice for the initial beliefs are their prior distributions. In the $n$-th iteration, the messages $m_{ij}^n(\alpha)$ and $m_{ij}^n(\mathbf{x}_i)$ are updated using \cref{alg:calculate message of alpha,alg:calculate message of position}, respectively, that will be given in \cref{sec:message update}. Then, the belief of each position, i.e., $B^n(\mathbf{x}_i), \; i \in \myset_u$, is updated, either using an importance sampler or using \cref{alg:update belief via auxiliary importance sampling}, to be given in \cref{sec:message multiplication}. Finally, the belief $B^n(\alpha)$ is updated, which will be discussed in \cref{sec:message multiplication} as well. These operations iterate until certain termination condition is met, for instance, when the maximal number of iterations $N_{\text{max}}$ is arrived.
Different from the existing works, this Bayesian framework treats both $\alpha$ and $\mathbf{x}_i, \; i \in \myset$, as random variables. It has the advantage that any prior knowledge on $\alpha$ and $\mathbf{x}_i, \; i \in \myset$, can be integrated. 
By doing so, $f(\alpha)$ reflecting the prior knowledge on any particular environment and $f(\mathbf{x}_i)$ representing the prior knowledge of any degree can be exploited. For instance, an anchor with imperfect position information can be easily handled in this framework.
Moreover, this framework provides marginal posterior estimate for each unknown parameter, which contains much more information than just one point estimate.

\begin{algorithm}
	\caption{Cooperative Localization Algorithms}
	\begin{algorithmic}[1]
		\STATE{\textbf{Initialization}: $B^0(\alpha)$ and $B^0(\mathbf{x}_i)$ for all $i \in \myset$ }
		\vspace{2mm}
		\STATE{\textbf{for} $n =1: N_{\text{max}}$}
		\vspace{1mm}
		\STATE{\hspace{2mm}\textbf{for} each $i \in \myset_u$}
% 		\vspace{1mm}
		\STATE{\hspace{4mm} \textbf{for} each $j \in \Gamma_i$}
		\STATE{\hspace{6mm} \textbf{if} $j>i$, \textbf{then} calculate $m^n_{ij}(\alpha)$, see Algorithm \ref{alg:calculate message of alpha}}
		\STATE{\hspace{6mm} compute $m^n_{ij}(\mathbf{x}_i)$, see Algorithm \ref{alg:calculate message of position}}
		\STATE{\hspace{4mm} \textbf{end for}}
% 		\vspace{1mm}
		\STATE{\hspace{4mm} update and broadcast $B^{n}(\mathbf{x}_i)$, see the importance \\ 
			   \hspace{4mm} sampler in \cref{sec:importance sampler} or \cref{alg:update belief via auxiliary importance sampling}}
		\STATE{\hspace{2mm} \textbf{end for}}
		\vspace{1mm}
		\STATE{\hspace{2mm} calculate $B^{n}(\alpha)$ using Eq.~\eqref{eq:belief update of alpha}}
		\vspace{1mm}
		\STATE{\textbf{end for}}
	\end{algorithmic}
	\label{alg:centralized}
\end{algorithm}

The main challenge in the proposed message passing algorithms lies in that there is no closed-form solution except for two special cases: case with discrete-valued variables and case with jointly Gaussian-distributed continuous-valued variables \cite{Sudderth2010}. 
In our problem, where the variables are continuous-valued but not jointly Gaussian distributed, we must resort to numerical approximation mechanisms.
One naive and simple numerical approximation scheme is to define a set of grid points, on which the beliefs and messages are evaluated.
There are two limitations in this approach. First, the number of the grid points grows exponentially with the dimensionality of the variable. Second, for a certain fixed granularity, the number of the grid points along one dimension grows linearly with its supported interval. Therefore, this approach is appropriate only when the variable is of low dimensionality and defined on a bounded interval, for instance, the $1$D PLE variable $\alpha$ varying in the range of $\left[1.5, 6\right]$ \cite{Rappaport2001}.
Alternatively, Monte-Carlo-based numerical approximation approaches have been proposed in \cite{Sudderth2003, Sudderth2010}, where both the beliefs and the messages are approximated based on a set of weighted samples. These samples are generated using certain stochastic methods, for instance, Markov Chain Monte Carlo methods in \cite{Sudderth2003,Sudderth2010}.  
These sample-based approaches provide an alternative to deal with high-dimensional variables or variables with infinite or relatively large support, such as the position variable $\mathbf{x}_i$, $i \in \myset_u$.
Taking all the above into consideration, we will discretize $\alpha$ and sample $\mathbf{x}_i$, $i \in \myset_u$, using stochastic sampling methods. More specifically, the messages and the belief of $\alpha$, e.g., $m_{ij}(\alpha)$ and $B(\alpha)$, are only evaluated on a set of predefined grid points $\left\lbrace \alpha_d^r\right\rbrace_{r = 1}^R$; while the messages and the beliefs of positions, e.g., $m_{ij}(\mathbf{x}_i)$ and $B(\mathbf{x}_i)$, are approximated based on weighted samples. 
In the next two sections, we will detail the approximation mechanisms for message updating and belief updating.

%%%%%%%%%%%%%%%%%%%%%%%%%%%%%%%%%%%%%%%%%%%%%%%%%%%%%%%%%%%%%%%%%%%%%%%%%%%%%%%%%%%%%%%%%%%%%%%%%%%%%%%%%%%
%+++Sample_Based_Message_Integration

\section{Updating Messages of Positions and PLE}
\label{sec:message update}
In this section, we consider how to update the messages $m_{ij}(\alpha)$ and $m_{ij}(\mathbf{x}_i)$ approximately. 
We proceed with the BP message update rule, and message updating using the SPAWN can be derived in analogy with the BP. For the moment, we assume that $\{ \mathbf{x}_i^{l, n-1}\}_{l = 1}^{L}$, $\{\mathbf{x}_j^{l, n-1}\}_{l = 1}^{L}$ and $\{B(\alpha_d^r)^{n-1}\}_{r = 1}^{R}$ are available, which are the equally weighted samples of $B^{n-1}(\mathbf{x}_i)$, those of $B^{n-1}(\mathbf{x}_j)$ and the evaluation values of $B^{n-1}(\alpha)$ at $\left\lbrace \alpha_d^r \right\rbrace_{r = 1}^{R} $, respectively. 

The message $m^n_{ij} (\alpha) $ can be updated by approximating the double integral in \cref{eq:message integration 2} using importance sampling \cite{Bishop2006}, giving rise to,
 \begin{align*}
  m^n_{ij} (\alpha) 				& \propto \sum^{L}_{l = 1} w^{l,n}_{ij \rightarrow \alpha} f(r_{ij} |  \mathbf{x}^{l}_i, \mathbf{x}^{l}_j, \alpha),  \\ 					
  w^{l, n}_{ij \rightarrow \alpha}  & \propto \frac{ B^{n-1}(\mathbf{x}^{l}_i)\cdot B^{n-1}(\mathbf{x}^{l}_j) }{ m^{n-1}_{ij}(\mathbf{x}^{l}_i) \cdot m^{n-1}_{ij}(\mathbf{x}^{l}_j) \cdot q(\mathbf{x}^{l}_i, \mathbf{x}^{l}_j) } ,  \end{align*}
where $\{ \mathbf{x}_i^l, \mathbf{x}_j^l \}_{l = 1}^{L}$ are samples from the proposal distribution $q(\mathbf{x}_i, \mathbf{x}_j)$, and $w^{l, n}_{ij \rightarrow \alpha} $ is the importance weight satisfying $\sum_{l = 1}^{L}w^{l, n}_{ij \rightarrow \alpha} = 1$. 
Based on the fact that $\mathbf{x}_i$ and $\mathbf{x}_j$ are decoupled in the non-normalized target distribution $\frac{B^{n-1}(\mathbf{x}_i)}{m_{ij}^{n-1}(\mathbf{x}_i)} \cdot \frac{B^{n-1}(\mathbf{x}_j)}{m_{ij}^{n-1}(\mathbf{x}_j)}$, we decouple $\mathbf{x}_i$ and $\mathbf{x}_j$ in the proposal distribution, resulting in $q(\mathbf{x}_i, \mathbf{x}_j) = q(\mathbf{x}_i)\cdot q(\mathbf{x}_j)$. 
The question that remains to answer is how to choose $q(\mathbf{x}_i)$ and $q(\mathbf{x}_j)$. A sensible choice for $q(\mathbf{x}_i)$ is the belief $B^{n-1}(\mathbf{x}_i)$. The reasons are twofold. First, the belief $B^{n-1}(\mathbf{x}_i)$ is part of $\mathbf{x}_i$'s non-normalized target distribution $B^{n-1}(\mathbf{x}_i)/m_{ij}^{n-1}(\mathbf{x}_i)$, and it approximates $f(\mathbf{x}_i\left| \mathbf{r} \right. )$. Second, the samples from $ B^{n-1}(\mathbf{x}_i) $ are available, and no extra effort is need. For the same reasons, $B^{n-1}(\mathbf{x}_j)$ is chosen as the proposal distribution $q(\mathbf{x}_j)$. Consequently, $m_{ij}^n(\alpha)$ is approximated to
\begin{subequations}
\label{eq:message integration alpha}
  \begin{align}
  m^n_{ij} (\alpha) 				& \propto \sum^{L}_{l = 1} w^{l,n}_{ij \rightarrow \alpha} f(r_{ij} |  \mathbf{x}^{l, n-1}_i, \mathbf{x}^{l, n-1}_j, \alpha) ,
									\label{eq:message integration 2 IS} \\ 
   w^{l, n}_{ij \rightarrow \alpha} & \propto \frac{1}{ m^{n-1}_{ij}(\mathbf{x}^{l,n-1}_i) \cdot m^{n-1}_{ij}(\mathbf{x}^{l,n-1}_j) } .
  \label{eq:message integration 2 weights}
  \end{align}
\end{subequations}
Here, $\{\mathbf{x}_i^{l,n-1}\}_{l = 1}^{L}$ and $\{\mathbf{x}_j^{l,n-1}\}_{l = 1}^{L}$ are samples from $B^{n-1}(\mathbf{x}_i)$ and $B^{n-1}(\mathbf{x}_j)$, respectively, and the importance weights fulfill $\sum_{l = 1}^{L} w^{l, n}_{ij \rightarrow \alpha} = 1.$ 
Since we have defined the grid points $\left\lbrace \alpha_d^r \right\rbrace_{r = 1}^{R}$, as a last step, $ m^n_{ij} (\alpha)$ is evaluated at $\left\lbrace \alpha_d^r \right\rbrace_{r = 1}^{R}$. As will be seen later in Section \ref{sec:message multiplication}, evaluating the messages of $\alpha$ at $\left\lbrace \alpha_d^r \right\rbrace_{r = 1}^{R}$ can facilitate updating the belief $B(\alpha)$ significantly. 

For $m^n_{ij}(\mathbf{x}_i)$ in \cref{eq:message integration 1}, we can directly combine the discretization approximation and the importance sampling technique, leading to 
\begin{subequations}
\label{eq:message integration 1 part IS}
 \begin{align}
	m_{ij}^n(\mathbf{x}_i) 				& \propto \sum_{r = 1}^{R} \sum^{L}_{l = 1} w^{l, n}_{ij \rightarrow \mathbf{x}_i} \frac{B^{n-1}(\alpha_d^r)}{m_{ij}^{n-1}(\alpha_d^r)} f( r_{ij} \big|  \mathbf{x}_i, \mathbf{x}^{l, n-1}_{j}, \alpha_d^{r} ) ,  \label{eq:message integration 1 complex}\\ 
	w^{l,n}_{ij\rightarrow \mathbf{x}_i}& \propto 1 / m^{n-1}_{ij}(\mathbf{x}^{l,n-1}_j),
\end{align}
\end{subequations}
where $\{ \mathbf{x}_j^{l, n-1} \}_{l = 1}^{L}$ denote the samples from $B^{n-1}(\mathbf{x}_j)$, and the importance weights fulfill $\sum_{l = 1}^L w^{l,n}_{ij\rightarrow \mathbf{x}_i} = 1$. 
In contrast to $m_{ij}^n(\alpha)$ in \cref{eq:message integration 2 IS}, the double-integral problem for $m_{ij}^n(\mathbf{x}_i)$ becomes a double summation in \cref{eq:message integration 1 complex}. As compared to $m_{ij}^n(\alpha)$, $m_{ij}^n(\mathbf{x}_i)$ is computationally heavier, incurring that updating the belief $B(\mathbf{x}_i)$ will be computationally intensive as well. In order to approximate $m_{ij}^n(\mathbf{x}_i)$ computationally more efficiently, we treat $\alpha$ in the same manner as $\mathbf{x}_j$ and perform importance sampling for both $\alpha$ and $\mathbf{x}_j$. 
However, $\alpha$ can be drawn only from the set of the grid points $\displaystyle\left\lbrace \alpha_d^r \right\rbrace_{r = 1}^{R}$, since its non-normalized target distribution $\displaystyle B^{n-1}(\alpha)/M^{n-1}_{ij}(\alpha)$ are evaluated at $\left\lbrace \alpha_d^r \right\rbrace_{r = 1}^{R}$. Subsequently, $m_{ij}^n(\mathbf{x}_i)$ can be approximated to
\begin{subequations}
%\label{eq:message integration 1 IS}
\begin{align}
   m_{ij}^n(\mathbf{x}_i) 								 & \propto \sum^{L}_{l = 1} w^{l, n}_{ij \rightarrow \mathbf{x}_i} f( r_{ij} \big|  \mathbf{x}_i, \mathbf{x}^{l, n-1}_{j}, \alpha^{l, n-1}) , \label{eq:message integration 1 IS} \\ 
	w^{l,n}_{ij\rightarrow \mathbf{x}_i} 				& \propto \frac{1}{ m^{n-1}_{ij}(\mathbf{x}^{l,n-1}_j) \cdot m^{n-1}_{ij}(\alpha^{l,n-1}) }  \label{eq:message integration 1 weights} ,
\end{align}
\end{subequations}
where $\{ \mathbf{x}_j^{l, n-1} \}_{l = 1}^{L}$ and $\{ \alpha^{l, n-1} \}_{l = 1}^{L}$ denote the samples from $B^{n-1}(\mathbf{x}_j)$ and $B^{n-1}(\alpha)$, respectively, and the importance weights fulfill $\sum_{l = 1}^L w^{l,n}_{ij\rightarrow \mathbf{x}_i} = 1$. Thank to the additional sampling process, the double summation in \cref{eq:message integration 1 complex} is simplified to a single summation in \cref{eq:message integration 1 IS}.

Next, we further transform the message $m_{ij}^n(\mathbf{x}_i)$ to 
\begin{subequations}
\label{eq:message after normalization}
	\begin{gather}
		m_{ij}^n(\mathbf{x}_i)  												  \propto \sum^{L}_{l = 1} \tilde{w}_{ij\rightarrow \mathbf{x}_i}^{l, n} \tilde{f}(r_{ij} \big| \mathbf{x}_i, \mathbf{x}^{l, n-1}_{j}, \alpha^{l, n-1})  \label{eq:message integration 1 IS new}, 
		\intertext{where $\tilde{f}(r_{ij} \big| \mathbf{x}_i, \mathbf{x}^{l, n-1}_{j}, \alpha^{l, n-1})$ is the normalized likelihood function, as given by}
		\tilde{f}(r_{ij} \big| \mathbf{x}_i, \mathbf{x}^{l, n-1}_{j}, \alpha^{l, n-1} )   =  Z_{ij}^{-1} f(r_{ij} \big| \mathbf{x}_i, \mathbf{x}^{l, n-1}_{j}, \alpha^{l, n-1} ), \label{eq:message conditional pdf after normalization} \\
		 Z_{ij} = \int f(r_{ij} \big| \mathbf{x}_i, \mathbf{x}^{l, n-1}_{j}, \alpha^{l, n-1} ) ~\mathrm{d}\mathbf{x}_i,
 \label{eq:normalization constant} 
		\intertext{and the mixture weight $\tilde{w}_{ij\rightarrow \mathbf{x}_i}^{l, n}$ is given by}
		\tilde{w}_{ij\rightarrow \mathbf{x}_i}^{l, n}	\propto   Z_{ij} \cdot w^{l,n}_{ij\rightarrow \mathbf{x}_i} , \label{eq:message weight after normalization}
\end{gather}
\end{subequations}
that satisfies $0 \leq \tilde{w}_{ij\rightarrow \mathbf{x}_i}^{l, n} \leq 1$ and $ \sum_{l = 1}^{L} \tilde{w}_{ij\rightarrow \mathbf{x}_i}^{l, n}  = 1 $. The integral in \cref{eq:normalization constant} can be evaluated analytically with the details given in Appendix \ref{sec:integral}. 
It is noteworthy that \cref{eq:message integration 1 IS new} differs from \cref{eq:message integration 1 IS} in that the mixture component $\tilde{f}(r_{ij} \left| \mathbf{x}_i, \mathbf{x}^{l}_{j}, \alpha^{l}\right.)$ is a normalized likelihood function of $\mathbf{x}_i$, satisfying the properties of a probability density function (pdf), while $f(r_{ij} | \mathbf{x}_i, \mathbf{x}^{l}_{j}, \alpha^{l})$ in \cref{eq:message integration 1 IS} not.
As will be seen later in Section \ref{sec:message multiplication}, $m_{ij}^n(\mathbf{x}_i)$ in the form of \cref{eq:message integration 1 IS new} is more advantageous than that in \cref{eq:message integration 1 IS}, since it enables the development of an efficient sampling procedure for updating $B(\mathbf{x}_i)$.
Finally, \cref{alg:calculate message of alpha,alg:calculate message of position} summarize the steps for updating the messages, $m_{ij}(\alpha)$ and $m_{ij}(\mathbf{x}_i)$, respectively.
 
\begin{remark}
In the SPAWN, the messages are updated in the same manner as the procedures above. The only difference is that the importance weights in \cref{eq:message integration 2 weights,eq:message integration 1 weights} are replaced by 
\begin{subequations}
\begin{align}
 w_{ij \rightarrow \alpha}^{l, n} & = 1/L \label{eq:message integration 2 weights (SPAWN)} ,\\ 
 w_{ij \rightarrow \mathbf{x}_i}^{l, n} & = 1/L \label{eq:message integration 1 weights (SPAWN)} .
\end{align}
\end{subequations}

\end{remark}

\begin{algorithm}
	\caption{Message Update of $m_{ij}^n(\alpha)$ }
	\begin{algorithmic}[1]
	 \STATE{ \textbf{Input}: $B^{n-1}(\mathbf{x}_i) \triangleq \{\mathbf{x}_i^{l, n-1}\}_{l = 1}^{L}$ \\[0.5mm] 
				\hspace{9mm} $B^{n-1}(\mathbf{x}_j) \triangleq \{\mathbf{x}_j^{l, n-1}\}_{l = 1}^{L}$}
	 \STATE{ \textbf{Output}: $\{ m_{ij}^{n}(\alpha_d^r) \}_{r = 1}^{R}$ }
	 \vspace{1mm}
	 \STATE{calculate $w_{ij\rightarrow \alpha}^{l, n}$ using Eq.~\eqref{eq:message integration 2 weights} $\leftarrow$ BP\\
         \hspace{2.8cm }\textbf{or} Eq.~\eqref{eq:message integration 2 weights (SPAWN)} $\leftarrow$ SPAWN} \\[0.8mm]
	 \STATE{evaluate $m_{ij}^n(\alpha)$ at $\left\lbrace \alpha_d^r \right\rbrace_{r = 1}^{R}$ using \cref{eq:message integration 2 IS}}
	\end{algorithmic}
	\label{alg:calculate message of alpha}
\end{algorithm}
\begin{algorithm}
	\caption{Message Update of $m_{ij}^{n}(\mathbf{x}_i)$ }
	\begin{algorithmic}[1]
	\STATE{\textbf{Input}: $B^{n-1}(\alpha)\triangleq \left\lbrace B^{n-1}(\alpha_d^r) \right\rbrace_{r = 1}^{R}$} \\ \qquad \quad 
	$B^{n-1}(\mathbf{x}_j) \triangleq \{\mathbf{x}_j^{l, n-1}\}_{l = 1}^{L}$
    \STATE{\textbf{Output}: $ \{ \tilde{w}_{ij\rightarrow \mathbf{x}_i}^{l,n}$, $\tilde{f}(r_{ij} \big| \mathbf{x}_i, \mathbf{x}^{l, n-1}_{j}, \alpha^{l, n-1} ) \}_{l  = 1}^{L}$}
	\vspace{2mm}
    \STATE{draw $\alpha^{l, n-1} \sim B^{n-1}(\alpha)$ }
    \STATE{calculate $w_{ij\rightarrow \mathbf{x}_i}^{l, n}$ using \cref{eq:message integration 1 weights} $\leftarrow$ BP\\
         \hspace{3cm }\textbf{or} \cref{eq:message integration 1 weights (SPAWN)} $\leftarrow$ SPAWN}
	\vspace{1mm}         
    \STATE{ compute $\tilde{w}_{ij\rightarrow \mathbf{x}_i}^{l,n}$ and $ \tilde{f}(r_{ij}  \big| \mathbf{x}_i, \mathbf{x}^{l, n-1}_{j}, \alpha^{l, n-1} )$ using \cref{eq:message after normalization}}
    \end{algorithmic}
	\label{alg:calculate message of position}
\end{algorithm}
%%%%%%%%%%%%%%%%%%%%%%%%%%%%%%%%%%%%%%%%%%%%%%%%%%%%%%%%%%%%%%%%%%%%%%%%%%%%%%%%%%%%%%%%%%%%%%%
%+++Sample_Based_Message_Multiplication+++
\section{Updating Beliefs of Positions And PLE}
\label{sec:message multiplication}

In this section, we will discuss the numerical approximation mechanism for updating the beliefs: $B(\alpha)$ and $B(\mathbf{x}_i)$, $i \in \myset_u$. First, we consider how to update the belief $B^n(\alpha)$ according to the update rule in \cref{eq:message multiplication 1}. For the reason that $\left\lbrace m^n_{ij}(\alpha_d^r) \right\rbrace_{r = 1}^{R}$ are available for each pair of connection $(i, j) \in \Gamma$, $B^n(\alpha)$ can be readily evaluated at $\left\lbrace \alpha_d^r \right\rbrace_{r = 1}^{R}$, 
\begin{equation}
 B^{n}(\alpha_d^r) \propto f_{\alpha}(\alpha_d^r) \prod_{(i,j) \in \Gamma} m^n_{ij}(\alpha_d^r).
 \label{eq:belief update of alpha}
\end{equation}
Thanks to the discretization, updating $B^n(\alpha)$ can be conducted by simply multiplying $|\Gamma|$ real-valued numbers at $R$ grid points. 

Next, we consider how to update the beliefs of position variables, for instance $B^n(\mathbf{x}_i)$. By combining \cref{eq:message multiplication 2,eq:message integration 1 IS new}, we obtain $B^{n}(\mathbf{x}_i)$ in the form of 
 \begin{equation}
%  \begin{aligned}
  \hspace{-0.3em} B^{n}(\mathbf{x}_i) 
                           \propto f(\mathbf{x}_i) \displaystyle{\prod_{j\in\Gamma_i}} \hspace{-1mm} \left( \sum_{l = 1}^{L} \tilde{w}_{ij\rightarrow \mathbf{x}_i}^{l, n} \tilde{f}(r_{ij} \big| \mathbf{x}_i , \mathbf{x}^{l, n-1}_{j}, \alpha^{l, n-1} ) \right) \hspace{-1mm}. 
 \label{eq:belief of position}
 \end{equation}
Our purpose is to conduct efficient sampling, i.e., $\mathbf{x}_i \sim B^n(\mathbf{x}_i)$. Here, the target distribution $B^n(\mathbf{x}_i)$ is a product of $\left| \Gamma_i \right|$ mixtures, each being a sum of $L$ weighted normalized likelihood functions. 
% \textcolor{red}{Since a normalized likelihood function fulfills the properties of a pdf, here we loosely refer to it as pdf.} 
Note that the component $\tilde{f}( r_{ij} \big| \mathbf{x}_i, \mathbf{x}^{l, n-1}_{j}, \alpha^{l, n-1} )$ is in general non-Gaussian. Therefore, updating $B(\mathbf{x}_i)$ boils down to sampling from a product of non-Gaussian mixtures. 
For notational convenience, we simplify Eq.~\eqref{eq:belief of position} to
\begin{align}
\label{eq:simplified belief of position}
  B(\mathbf{x}) \propto f(\mathbf{x})\displaystyle{\prod_{j = 1}^{J}} M_{j}(\mathbf{x}), \quad M_{j}(\mathbf{x}) & = \sum_{l = 1}^{L} \nu_{j}^l f_{j}^l(\mathbf{x}) .
\end{align}
One straightforward sampling strategy is to construct all components explicitly and to sample from them. This is, however, computationally prohibitive, since the product of $J$ mixtures, each containing $L$ components, is itself a mixture of $L^J$ components.
Besides, there exist several samplers in the existing works, including the Gibbs sampler \cite{Sudderth2003} and its related multi-scale sampling strategies in \cite{Ihler2003,Rudoy2007}. These approaches, however, require a prerequisite that each $M_j(\mathbf{x})$ is a Gaussian mixture, and, therefore, they are not applicable to our problem. 
In what follows, we will first revisit an existing sampling approach and then propose an alternative sampler, which has a significantly reduced computational complexity. 

\subsection{Importance Sampling as Baseline}
\label{sec:importance sampler}
First, we consider the technique of importance sampling. 
The samples and the associated weights are obtained as follows:
\begin{equation}
	\mathbf{x}^l \sim q(\mathbf{x}), \quad \quad 	w^l \propto B(\mathbf{x}^l) / q(\mathbf{x}^l),
	\label{eq:importance sampler}
\end{equation}
where $q(\mathbf{x})$ is an appropriate proposal distribution, and the importance weight $w^l$ satisfies $\sum_{l = 1}^L w^l = 1$. The possible choices for $q(\mathbf{x})$ are the prior distribution $f(\mathbf{x})$, an evenly weighted sum of $J$ mixtures $\sum_{j = 1}^{J} J^{-1} M_j(\mathbf{x})$ \cite{Ihler2005} and the message with the smallest entropy, e.g., $M_j(\mathbf{x})$, \cite{Lien2012}. 
The resulting \cref{alg:centralized} with the beliefs updated using the importance sampler in \cref{eq:importance sampler} is referred to as BP-IS or SPAWN-IS, for that the messages are updated according to the BP or the SPAWN, respectively. One shortcoming of the importance sampler lies in the high computational load, since computing these $L$ weights $\left\lbrace w \right\rbrace_{l = 1}^L$ according to \cref{eq:importance sampler} requires operations of order $\mathcal{O}(JL^2)$ \cite{Lien2012}. In order to reduce the computational load, we propose an alternative sampler in what follows.

\subsection{Proposed Auxiliary Importance Sampler}
\label{sec:AIS}
Motivated by \cite{Briers05}, we develop an efficient sampler, named as \emph{auxiliary importance sampler} (AIS), for the sampling problem $\mathbf{x} \sim B(\mathbf{x})$.
The key idea is to introduce an auxiliary variable $\psi_j$ to each mixture $M_j(\mathbf{x})$. The auxiliary variable $\psi_{j}$ plays the role of a component label indicator, indicating which component is drawn from the mixture $M_j(\mathbf{x}) = \sum_{l = 1}^{L} \nu_{j}^l f_{j}^l(\mathbf{x})$, and it can take value $\psi_{j} = \kappa$, where $\kappa \in \left\lbrace 1, \ldots, L \right\rbrace.$ For instance, if we have $\psi_{j} = \kappa$, it denotes that the $\kappa$-th component $\nu_{j}^{\kappa} f_{j}^{\kappa}(\mathbf{x})$ is drawn from the mixture $M_j(\mathbf{x}) = \sum_{l = 1}^{L} \nu_{j}^l f_{j}^l(\mathbf{x})$.
Stacking all $J$ auxiliary variables into a vector, we have the compact auxiliary variable $\boldsymbol{\psi} = \left[ \psi_1, \ldots, \psi_J\right]^T$.

With the help of the auxiliary variable $\boldsymbol{\psi}$, the sampling task $\mathbf{x} \sim B(\mathbf{x})$ can be achieved in two steps:
\begin{enumerate}
 \item Draw $\boldsymbol{\psi}^l \sim p(\boldsymbol{\psi} )$, \vspace{-3mm}
	\begin{align}
	  \hspace{-5mm} p(\boldsymbol{\psi} ) = \int f(\mathbf{x}, \boldsymbol{\psi}) \mathrm{d}\mathbf{x} = Z_1^{-1} \int \prod_{j = 1}^{J} \nu_j^{\psi_j} f_{j}^{\psi_j}(\mathbf{x}) ~\mathrm{d}\mathbf{x} ; \label{eq:pdf of component label}
	\end{align} 
%	\vspace{-3mm}
%  
 \item Draw $\mathbf{x}^l \sim f(\mathbf{x} | \boldsymbol{\psi}^l)$, conditional on $\boldsymbol{\psi}^l$, 
% \vspace{-2mm}
 \begin{align}
  f(\mathbf{x} | \boldsymbol{\psi}^l) = Z_2^{-1} \prod_{j = 1}^{J} f_{j}^{\psi_j^l}(\mathbf{x}) .	 \label{eq:pdf of location conditional on component label}  
 \end{align}
%  \vspace{-5mm}
\end{enumerate}
Here, $Z_1$ and $Z_2$ are two normalization constants. Neglecting the auxiliary variable samples $\{ \boldsymbol{\psi}^{l} \}_{l = 1}^{L}$, the samples $\{ \mathbf{x}^{l} \}_{l = 1}^{L}$ generated in such a two-step procedure follow the distribution in Eq. \eqref{eq:simplified belief of position}. 
However, when directly sampling from $p(\boldsymbol{\psi} )$ and $f(\mathbf{x} | \boldsymbol{\psi}^l)$ is impossible, as in our case, we can generate samples from two proposal distributions $ q(\boldsymbol{\psi}) $ and $ q(\mathbf{x} | \boldsymbol{\psi}^l) $ and assign certain importance weights to them. This gives rise to the following three-step procedure:
\begin{enumerate}
 \item Draw $ \boldsymbol{\psi}^l \sim q(\boldsymbol{\psi} ) $; 
 \item Draw $\mathbf{x}^l \sim q(\mathbf{x} | \boldsymbol{\psi}^l)$, conditional on $\boldsymbol{\psi}^l$;
 \item Calculate the importance weight $w^l$
 \begin{align*}
 	 w^l\propto \frac{ f(\mathbf{x}^l,\; \boldsymbol{\psi}^{l}) }{q(\mathbf{x}^l, \boldsymbol{\psi}^l)} = \frac{ f(\mathbf{x}^l,\; \boldsymbol{\psi}^{l}) }{q(\boldsymbol{\psi}^l)\cdot q(\mathbf{x}^l \left| \boldsymbol{\psi}^l \right.)}
 \end{align*} 
 with the non-normalized joint distribution $f(\mathbf{x}^l, \boldsymbol{\psi}^{l})$ given by
\begin{align*}
	 f(\mathbf{x}^l, \boldsymbol{\psi}^{l}) = f(\mathbf{x}^l) \prod_{j = 1}^{J} \nu_j^{\psi_j^l} f_j^{\psi_j^l}(\mathbf{x}^l) . 
	 \end{align*}
\end{enumerate}
% \vspace{-3mm}
% 
Up to this point, the problem remained is how to design $q(\boldsymbol{\psi})$ and $q(\mathbf{x} | \boldsymbol{\psi}^l)$, which will be addressed in what follows. 

\begin{remark}
Note that the underlying condition in the AIS is that the target distribution $B(\mathbf{x})$ must be a product of several mixtures, each being a sum of multiple weighted pdfs. Thanks to the additional message transformation in \cref{eq:message after normalization}, the belief in \cref{eq:belief of position} satisfies the properties of this condition, meaning that the message transformation in \cref{eq:message after normalization} is a prerequisite for the development of the AIS.
\end{remark}

\subsubsection{Auxiliary Variable $\boldsymbol{\psi}$}
\label{sec:auxiliary variable}
First, we focus on designing an appropriate proposal distribution $q(\boldsymbol{\psi})$. Ideally, $q(\boldsymbol{\psi})$ should resemble the corresponding target distribution $p(\boldsymbol{\psi})$ as closely as possible, and, at the same time, it should be feasible to draw samples from it. To this end, we first recover the original form of the target distribution $p(\boldsymbol{\psi})$. 
This can be readily achieved by replacing $\mathbf{x}$, $\nu_j^{\psi_j}$ and $f_j^{\psi_j}(\mathbf{x})$ in \cref{eq:pdf of component label} with $\mathbf{x}_i$, $\tilde{w}_{ij \rightarrow \mathbf{x}_i}^{\psi_j} $ and $ \tilde{f}(r_{ij}\big| \mathbf{x}_i, \mathbf{x}_j^{\psi_j}, \alpha^{\psi_j} ) $, respectively, giving rise to 
\begin{equation*}
 p(\boldsymbol{\psi}) \propto \int \prod_{j\in \Gamma_i} \tilde{w}_{ij \rightarrow \mathbf{x}_i}^{\psi_j} \tilde{f}(r_{ij} \big| \mathbf{x}_i, \mathbf{x}_j^{\psi_j}, \alpha^{\psi_j} ) \; \mathrm{d} \mathbf{x}_i .
\end{equation*}
To ensure mathematical tractability, we assume that all auxiliary variables in $\left\lbrace \psi_j : j \in \Gamma_i\right\rbrace$ are independent, giving rise to $\displaystyle q(\boldsymbol{\psi}) = \prod_{j = \Gamma_i} q(\psi_{j})$ with
 \begin{align*}
 q(\psi_{j} \! =  \!\kappa)  = \tilde{w}_{ij \rightarrow \mathbf{x}_i}^{\kappa}  \int  \tilde{f}(r_{ij} \left| \mathbf{x}_i, \mathbf{x}_j^{\kappa}, \alpha^{\kappa} \right. ) \; \mathrm{d} \mathbf{x}_i   =  \tilde{w}_{ij \rightarrow \mathbf{x}_i}^{\kappa}  , \label{eq:definition of proposal of auxiliary variable}
\end{align*}
where the second equality follows from \cref{eq:message conditional pdf after normalization,eq:normalization constant}. 

\subsubsection{Position Variable $\mathbf{x}$}
In order to design $q(\mathbf{x} | \boldsymbol{\psi}^l)$, again, we recover the original form of $f(\mathbf{x} | \boldsymbol{\psi}^l)$. This is done by replacing $\mathbf{x}$ and $f_j^{\psi_j^l}(\mathbf{x})$ in \cref{eq:pdf of location conditional on component label} with $\mathbf{x}_i$ and $\tilde{f}(r_{ij}| \mathbf{x}_i, \mathbf{x}_j^{\psi_j^l}, \alpha^{\psi_j^l})$, respectively, giving rise to
\begin{equation}
 f(\mathbf{x}_i | \boldsymbol{\psi}^l)  \propto \prod_{j \in \Gamma_i} \tilde{f}(r_{ij} \big| \mathbf{x}_i, \mathbf{x}_j^{\psi_j^l}, \alpha^{\psi_j^l} ) .
\label{eq:pdf conditional on a component label}
\end{equation}
To capture each mixture component in Eq.~\eqref{eq:pdf conditional on a component label}, we design $q(\mathbf{x}_i | \boldsymbol{\psi}^l)$ in the form of
\begin{align}
q(\mathbf{x}_i | \boldsymbol{\psi}^l) =
%\sum_{j\in\Gamma_i} |\Gamma_i|^{-1}q(\mathbf{x}_i | \psi_j^l),
\sum_{j\in\Gamma_i} |\Gamma_i|^{-1}q(\mathbf{x}_i \big| \mathbf{x}_j^{\psi_j^l}, \alpha^{\psi_j^l}, r_{ij}),
\label{eq:proposal of position}
\end{align}
where $q(\mathbf{x}_i |\mathbf{x}_j^{\psi_j^l}, \alpha^{\psi_j^l}, r_{ij})$ should resemble $ \tilde{f}(r_{ij}| \mathbf{x}_i, \mathbf{x}_j^{\psi_j^l}, \alpha^{\psi_j^l})$ as closely as possible, and, at the same time, drawing samples from it remains feasible. 
For notational clarity, we will replace $\psi_j^l$ with $l'$, thereby simplifying $q(\mathbf{x}_i |\mathbf{x}_j^{\psi_j^l}, \alpha^{\psi_j^l}, r_{ij})$ to $q(\mathbf{x}_i |\mathbf{x}_j^{l'}, \alpha^{l'}, r_{ij})$. 

% \textbf{Sampling from a normalized likelihood function}: 
Next, we proceed with designing the proposal distribution $q(\mathbf{x}_i |\mathbf{x}_j^{l'}, \alpha^{l'}, r_{ij})$ for the target distribution $ \tilde{f}(r_{ij}| \mathbf{x}_i, \mathbf{x}_j^{l'}, \alpha^{l'})$, which is the normalized likelihood function $f(r_{ij} | \mathbf{x}_i, \mathbf{x}_j^{l'}, \alpha^{l'} )/\int f(r_{ij} | \mathbf{x}_i, \mathbf{x}_j^{l'}, \alpha^{l'} ) \mathrm{d} \mathbf{x}_i$. This task is actually an embedded step in many other works, for instance, under different measurement models in \cite{Ihler2005,Wymeersch2009, Lien2012, Jin2015, Jin2016}. Therefore, instead of being specific, we generalize this sampling problem to a generic measurement model, given by
\begin{align}
 r_{ij} = h(d_{ij}) + v, \quad v  \sim f_{v}(v) .
 \label{eq:measurement model simplified}
\end{align}
Here $r_{ij}$ denotes any distance-related measurement, $h(d_{ij})$ is a function of the internode distance $d_{ij} = \lVert \mathbf{x}_i - \mathbf{x}_j \rVert$, and $v$ is an additive measurement error. 
Our purpose is to sample from the normalized likelihood function, namely, 
% 
% \begin{subequations}
\begin{align}
\mathbf{x}_i^l & \sim Z^{-1}f(r_{ij} \big| \mathbf{x}_i, \mathbf{x}_j^{l'} ) ,
\label{eq:sampling position for simplified measurement model}
\end{align}
% \end{subequations}
% 
where $\mathbf{x}_j^{l'}$ is a reference position, and $Z$ is a normalization constant, to be precise, $Z =   \int f(r_{ij} \big| \mathbf{x}_i, \mathbf{x}_j^{l'} ) \; \mathrm{d}\mathbf{x}_i $. The proposal distribution $q(\mathbf{x}_i \big| \mathbf{x}_j^{l'}, r_{ij} )$ for the sampling problem in \cref{eq:sampling position for simplified measurement model} can be designed in a bottom-up manner, meaning that we first develop a sampling strategy and then derive the associated distribution $q(\mathbf{x}_i \big| \mathbf{x}_j^{l'}, r_{ij} )$. Given $r_{ij}$, $\mathbf{x}_j^{l'}$ and the measurement model in \cref{eq:measurement model simplified}, an intuitive and reasonable approach to generate $\mathbf{x}_i^l$ is as follows:
\begin{subequations}
\label{eq:position samples in existing work}
 \begin{align}
\theta_{ij}^l       & \sim \mathcal{U}\left[ 0, 2\pi\right), \label{eq:angle sample}\\
 v^l          & \sim f_{v}(v), \label{eq:noise sample} \\
 d_{ij}^l            & = h^{-1}\left( r_{ij} - v^l \right), \label{eq:distance sample}\\ 
 \mathbf{x}_i^l & = \mathbf{x}_j^{l'} +  \left[d_{ij}^l \cdot \cos \theta_{ij}^l, \; d_{ij}^l \cdot \sin \theta_{ij}^l \right]^T \label{eq:position sample}.
\end{align}
\end{subequations}
In words, the sample $\mathbf{x}_i^l$ is obtained by moving $\mathbf{x}_j^{l'}$ in a random direction $\theta_{ij}^l$ by a random distance $d_{ij}^l$, which is generated based on the measurement model and the measurement $r_{ij}$. 
We denote the distributions of $\theta_{ij}$, $d_{ij}$ and $\mathbf{x}_i$ associated with the sampling procedures in \cref{eq:angle sample,eq:distance sample,eq:position sample} by $q_{\theta}(\theta_{ij})$, $q_d(d_{ij} | r_{ij})$ and $q(\mathbf{x}_i | \mathbf{x}_j^{l'}, r_{ij})$, respectively. Note that the subscripts $\theta$ and $d$ are introduced in $q_{\theta}(\theta_{ij})$ and $q_d(d_{ij} | r_{ij})$ to indicate the distributions of $\theta_{ij}$ and $d_{ij}$, respectively.
However, it seems not straightforward to obtain the proposal distribution $q(\mathbf{x}_i \big| \mathbf{x}_j^{l'}, r_{ij} )$. 

As one of our contributions, we provide a mathematical interpretation and justification for the sampling procedure in \cref{eq:position samples in existing work}, upon which, we further derive the proposal distribution $q(\mathbf{x}_i \big| \mathbf{x}_j^{l'}, r_{ij} )$.
The underlying idea of the sampling procedure in \cref{eq:position samples in existing work} is the transformation between a pair of random variables, from polar coordinate $\left[d_{ij}, \theta_{ij}\right]^T$ to Cartesian coordinate $\mathbf{x}_i $. Equivalently speaking, drawing the position sample $\mathbf{x}_i^l$ is transformed to a problem of drawing the sample pair of distance and angle, i.e., $\left[ d_{ij}^l, \theta_{ij}^l \right]^T $. As a consequence, the distributions $q(\mathbf{x}_i \big| \mathbf{x}_j^{l'}, r_{ij} )$ and $q_{d, \theta}(d_{ij}, \theta_{ij} | r_{ij}) = q_{d}(d_{ij} \left| r_{ij} \right. )\cdot q_{\theta}(\theta_{ij})$ are related according to
\begin{align}
  q(\mathbf{x}_i | \mathbf{x}_j^{l'}, r_{ij}) & = \frac{ q_{d}\left( d_{ij} = \lVert \mathbf{x}_i - \mathbf{x}_j^{l'}\rVert \big| r_{ij}  \right)  }{2\pi \cdot \lVert \mathbf{x}_i - \mathbf{x}_j^{l'}\rVert } .
 \label{eq:proposal distribution using variable transform}
\end{align}
Thanks to \cref{eq:proposal distribution using variable transform}, deriving $q(\mathbf{x}_i \big| \mathbf{x}_j^{l'}, r_{ij} )$ reverts to the problem of deriving $q_{d}(d_{ij} | r_{ij} )$, which should not be difficult for most measurement models. In our problem, where the measurement model is defined in \cref{eq:measurement model}, $q_d(d_{ij} \left| r_{ij} \right.)$ is derived as
\begin{subequations}
\label{eq:proposal distribution}
 \begin{align}
	\hspace{-3mm} q_d(d_{ij} \left|  r_{ij} \right.)               & = \frac{1}{\sqrt{2\pi} \frac{d_{ij}}{d_0} \tilde{\sigma}} \text{exp}\left(- \frac{1}{2 \tilde{\sigma}^2} \left(\text{log}\frac{d_{ij}}{d_0}-\tilde{\mu} \right)^2 \right),  \label{eq:conditional pdf of distance} \\ 
	\tilde{\mu}                                 & =  \frac{\text{log}10}{10\alpha^{l'}} \cdot \left( A-r_{ij} \right), \\
	 \tilde{\sigma}^2                    & =  \left(\frac{\text{log}10}{10\alpha^{l'}}\right)^2 \hspace{-2mm} \cdot \sigma^2 .  
%   \label{eq:relation between angles and positions} 
\end{align}
\end{subequations}
Replacing $q_d(d_{ij}\left| r_{ij}\right.)$ in \cref{eq:proposal distribution using variable transform} with \cref{eq:proposal distribution} gives rise to the proposal distribution  $q(\mathbf{x}_i | \mathbf{x}_j^{l'}, r_{ij})$, which is equivalent to $q( \mathbf{x}_i \big| \mathbf{x}_j^{l'}, \alpha^{l'} , r_{ij} ) $ in our original problem. 
More details about Eq.~\eqref{eq:proposal distribution} are given in Appendix \ref{sec:conditional pdf of distance}. 

\subsubsection{Importance Weight $w_i^l$}
For the auxiliary variable sample $\boldsymbol{\psi}^l$ and the position sample $\mathbf{x}_i^l$, which are generated from $q(\boldsymbol{\psi})$ and $q(\mathbf{x}_i | \boldsymbol{\psi}^l)$, respectively, the associated importance weight $w_i^l$ is given by 
\begin{align}
 w_i^l \propto \frac{ \prod_{j \in \Gamma_i} f( r_{ij} \big| \mathbf{x}_i^l, \mathbf{x}_j^{\psi_j^l}, \alpha^{\psi_j^l} ) }{ \sum_{j\in\Gamma_i} |\Gamma_i|^{-1}q(\mathbf{x}_i^l \big| \mathbf{x}_j^{\psi_j^l}, \alpha^{\psi_j^l}, r_{ij}) }.
 \label{eq:weight in AIS}
\end{align}

Finally, Algorithm \ref{alg:update belief via auxiliary importance sampling} lists the steps for updating $B(\mathbf{x}_i)$ using the proposed AIS. The resulting \cref{alg:centralized} with the beliefs updated using \cref{alg:update belief via auxiliary importance sampling} are named as BP-AIS or SPAWN-AIS, for that the messages are updated according to the BP or the SPAWN, respectively.
\begin{algorithm}
	\caption{Belief Update Using AIS}
	\begin{algorithmic}[1]
    \STATE{\textbf{Input}: $m^n_{ij}(\mathbf{x}_i)$ for all $j\in\Gamma_i$}
    \vspace{2mm}
    \STATE{\textbf{Output}: $B^{n}(\mathbf{x}_i) \triangleq \{ \mathbf{x}_i^{l, n} \}_{l = 1}^{L}$}
	\vspace{2mm}
	\STATE{draw $\boldsymbol{\psi}^l \sim q(\boldsymbol{\psi})$ as follows:}
		\STATE{\hspace{4mm}\textbf{for} each $j \in \Gamma_i$ }
			\STATE{\hspace{8mm} draw $\psi_j^l \sim q(\psi_j)$}
		\STATE{\hspace{4mm}\textbf{end for}}
    \vspace{2mm}
	\STATE{draw $\mathbf{x}_i^l \sim q(\mathbf{x}_i \left| \boldsymbol{\psi}^l\right.)$ as follows:}
		\STATE{\hspace{4mm} \textbf{for} each $j \in \Gamma_i$}
			\STATE{\hspace{8mm} draw $\mathbf{x}_i^l \sim q(\mathbf{x}_i | \mathbf{x}_j^{\psi_j^l}, \alpha^{\psi_j^l}, r_{ij} )$ using Eq.~\eqref{eq:position samples in existing work}}
% 			\COMMENT{This step is implemented using Eq.~\eqref{eq:position sample all}}
		\STATE \textbf{\hspace{4mm} end for}
    \vspace{2mm}
	\STATE calculate $w_i^l$ using Eqs. \eqref{eq:proposal distribution using variable transform}-\eqref{eq:weight in AIS}. 
	\STATE resampling 
% 	$\rightarrow \mathbf{x}_i^{\{l\}, n}$ 
% 	\STATE resampling and obtain $\mathbf{x}_i^{\{l\}, n+1}$
    \end{algorithmic}
	\label{alg:update belief via auxiliary importance sampling}
\end{algorithm}

%%%%%%%%%%%%%%%%%%%%%%%%%%%%%%%%%%%%%%%%%%%%%%%%%%%%%%%%%%%%%%%%%%%%%%%%%%%%%%%%%%%%%%%%%%%%%%
%+++Some_Other_Issues+++

\section{Some Important Issues}
\label{sec:other issues}
\subsection{Computational Complexity}
\label{sec:computational complexity}
{
\renewcommand{\arraystretch}{1.5}
\begin{table}[ht!]
\centering
	\begin{tabular}{ c | l | l | l }
		\hline \hline
	\multirow{3}{*}{$m_{ij}(\alpha)$}	&	\multirow{2}{*}{importance weight}          & $\mathcal{O}(L^2)$ & \cref{eq:message integration 2 weights} in BP \\ \cline{3-4}
											&												& $\mathcal{O}(1)$ & \cref{eq:message integration 2 weights (SPAWN)} in SPAWN \\ \cline{2-4}  
											&	evaluating	$m_{ij}(\alpha_d^r)$			& \multicolumn{2}{l}{$ \mathcal{O} ( L\cdot R )$ } \\ \hline \hline
	\multirow{4}{*}{$m_{ij}(\mathbf{x}_i)$} & sample $\alpha$								& \multicolumn{2}{l}{$ \mathcal{O} ( \mathbb{C}_c(L, R))$ } \\ 
												\cline{2-4}
											&	\multirow{2}{*}{importance weight}          & $\mathcal{O}(L^2)$ & \cref{eq:message integration 1 weights} in BP \\  \cline{3-4}
											   &											& $\mathcal{O}(1)$ &  \cref{eq:message integration 1 weights (SPAWN)} in SPAWN \\  \cline{2-4}
											   &normalization								& \multicolumn{2}{l}{$ \mathcal{O} ( L)$ }\\ 
		\hline \hline
	\end{tabular}
\caption{\small{Complexity of updating $m_{ij}(\alpha)$ and $m_{ij}(\mathbf{x}_i)$}}
\label{tab:complexity of message update for alpha}
\end{table}
}
{
\renewcommand{\arraystretch}{1.5}
\begin{table}[ht!]
\centering
	\begin{tabular}{ L{0.7cm} | L{1.5cm} | L{2.3cm} | L{2.5cm} }
		\hline \hline
		\multirow{7}{*}{$B(\mathbf{x}_i)$} &
		\multirow{3}{*}{\parbox{1.5cm}{importance sampler} }                
								& sampling   & $\mathcal{O}(L)$  \\ \cline{3-4}
								&& \textbf{importance weight} & $\boldsymbol{\mathcal{O}(|\Gamma_i| \cdot L^2)}$  \\ \cline{3-4} 
								&& resampling        & $ \mathcal{O} ( L)$\\  \cline{2-4}
&	\multirow{4}{*}{\parbox{1.5cm}{auxiliary importance sampling}}	
								& label indicator   & $\mathcal{O}(|\Gamma_i| \cdot \mathbb{C}_{\text{c}}(\frac{L}{|\Gamma_i|}, L) )$  \\ \cline{3-4}
&								& position sample   & $\mathcal{O}(L)$ \\ \cline{3-4} 
&								& \textbf{importance weight} & $\boldsymbol{\mathcal{O}(|\Gamma_i| \cdot L)}$ \\ \cline{3-4}
&		                        & resampling        & $ \mathcal{O} ( L )$\\  \hline\hline
	\multicolumn{2}{l|}{$B(\alpha)$}				& evaluate $B(\alpha_d^r)$ & $\mathcal{O}(|\Gamma|\cdot R)$ \\ \hline \hline
	\end{tabular}
\caption{\small{Complexity of updating $B(\mathbf{x}_i)$ and $B(\alpha)$. Here, the bold fonts are used to highlight the reduction in computational complexity, where the quadratic order in the importance sampler is reduced to the linear order in the proposed AIS.}}
\label{tab:complexity of belief update}
\end{table}
}
\vspace{0mm}
In this subsection, the four main parts of \cref{alg:centralized}, including updating $m_{ij}(\alpha)$, $m_{ij}(\mathbf{x}_i)$, $B(\mathbf{x}_i)$ and $B(\alpha)$, will be analyzed in terms of computational complexity. To be general, we write $\mathbb{C}_{\text{c}}(M, N)$ to denote the complexity of drawing $M$ samples from an $N$-categorical distribution. 
First, we consider updating $m_{ij}(\alpha)$ using \cref{alg:calculate message of alpha}. Importance weights $\{ w_{ij \rightarrow \alpha}^{l, n} \}_{l = 1}^{L}$ are calculated with a complexity order of $\mathcal{O}(L^2)$ according to \cref{eq:message integration 2 weights} in the BP, but $\mathcal{O}(1)$ according to \cref{eq:message integration 2 weights (SPAWN)} in the SPAWN. Evaluating $m_{ij}(\alpha)$ at $\{\alpha_d^r \}_{r = 1}^{R}$ requires operations of order $\mathcal{O}(L\cdot R)$.
Second, for updating $m_{ij}(\mathbf{x}_i)$ using \cref{alg:calculate message of position}, drawing samples $\{ \alpha^{l} \}_{l = 1}^{L}$ from $B(\alpha)$ needs operations of order $\mathcal{O}(\mathbb{C}_{\text{c}}(L, R))$, calculating $\{ w_{ij \rightarrow \mathbf{x}_i}^{l, n} \}_{l = 1}^{L}$ has the same complexity as for $\{ w_{ij \rightarrow \alpha}^{l, n} \}_{l = 1}^{L}$, and converting $m_{ij}(\mathbf{x}_i)$ from \cref{eq:message integration 1 IS} to \cref{eq:message integration 1 IS new} is done with a complexity order of $\mathcal{O}(L)$.
Third, $B(\mathbf{x}_i)$ can be updated either using the importance sampler or using the proposed AIS. 
For the importance sampler in \cref{eq:importance sampler}, $L$ position samples and the corresponding importance weights are obtained with complexity orders of $\mathcal{O}(L)$ and $\mathcal{O}(|\Gamma_i| \cdot L^2)$, respectively. The subsequent resampling is conducted with a complexity order of $\mathcal{O}(L)$ \cite{Gustafsson2006}.
For the proposed AIS in \cref{alg:update belief via auxiliary importance sampling}, generating $L$ label indicators has a complexity order of $\mathcal{O}(|\Gamma_i| \cdot \mathbb{C}_{\text{c}}(\frac{L}{|\Gamma_i|}, L) )$ approximately. Generating $L$ position samples and calculating $L$ importance weights according to \cref{eq:weight in AIS} have complexity orders of $\mathcal{O}(L)$ and $\mathcal{O}(|\Gamma_i|\cdot L)$, respectively. The resampling step requires additional operations of order $\mathcal{O}(L)$.
Lastly, $B(\alpha)$ is updated by simply multiplying $|\Gamma|$ real-valued numbers for $R$ times, according to \cref{eq:belief update of alpha}. 

\begin{figure*}[ht!]
\begin{subfigure}{0.33\textwidth}
\centering
 \includegraphics[width = 1.0 \textwidth]{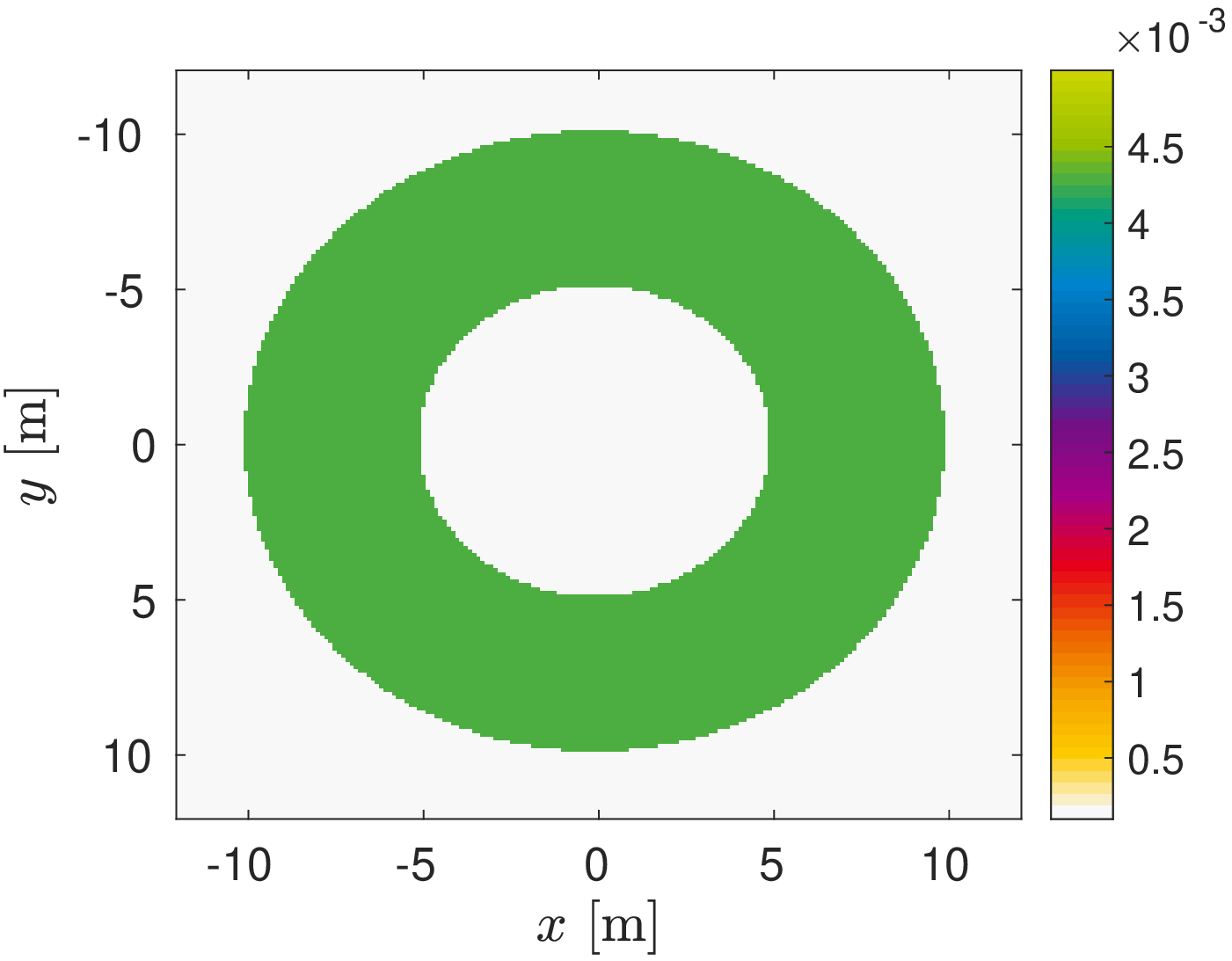}
%  \caption{}
 \label{fig:groundtruth_broad}
\end{subfigure}
\begin{subfigure}{0.33\textwidth}
\centering
 \includegraphics[width = 1.0 \textwidth]{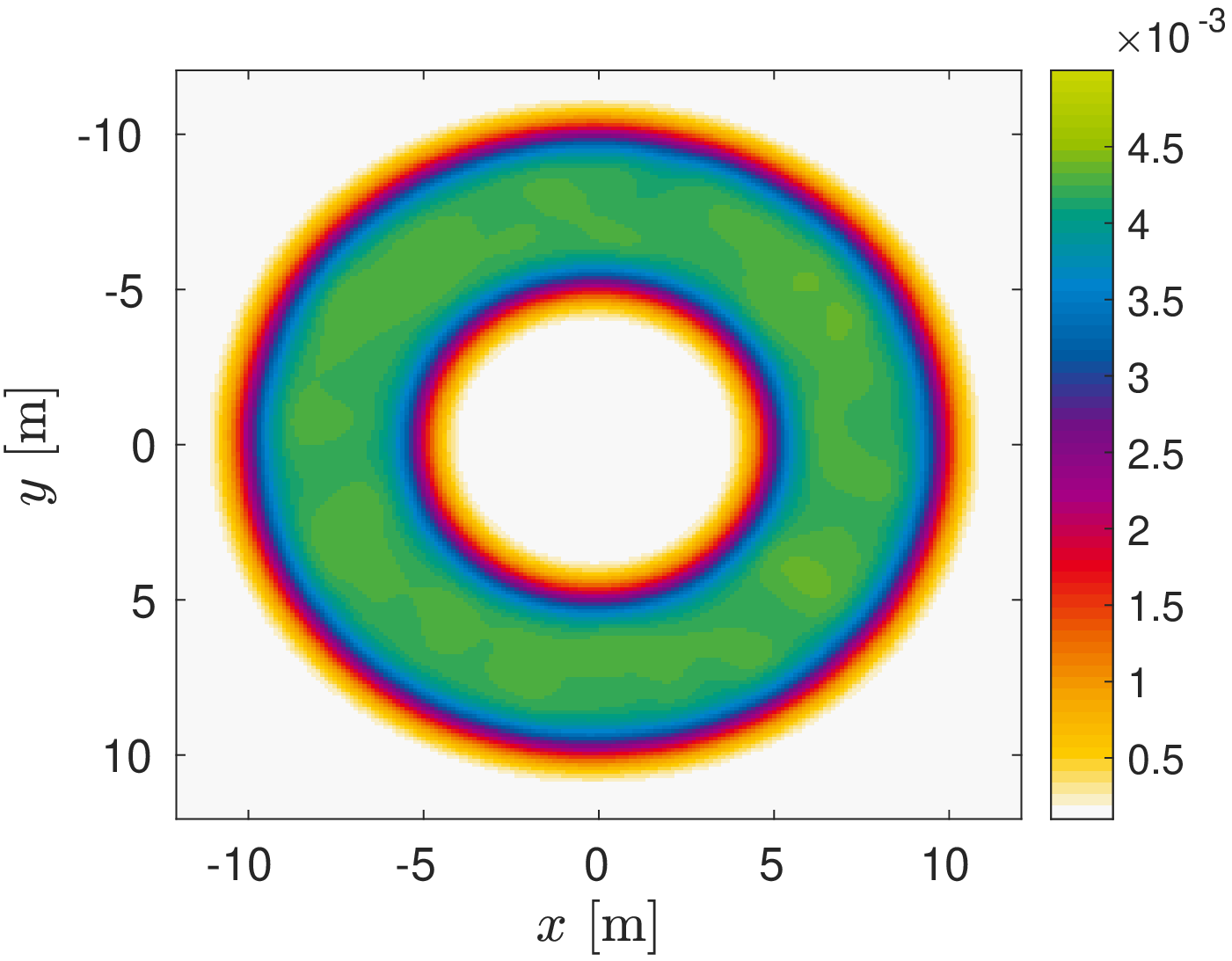}
%  \caption{}
 \label{fig:ISVT sampler_broad}
\end{subfigure}
\begin{subfigure}{0.33\textwidth}
\centering
 \includegraphics[width = 1.0 \textwidth]{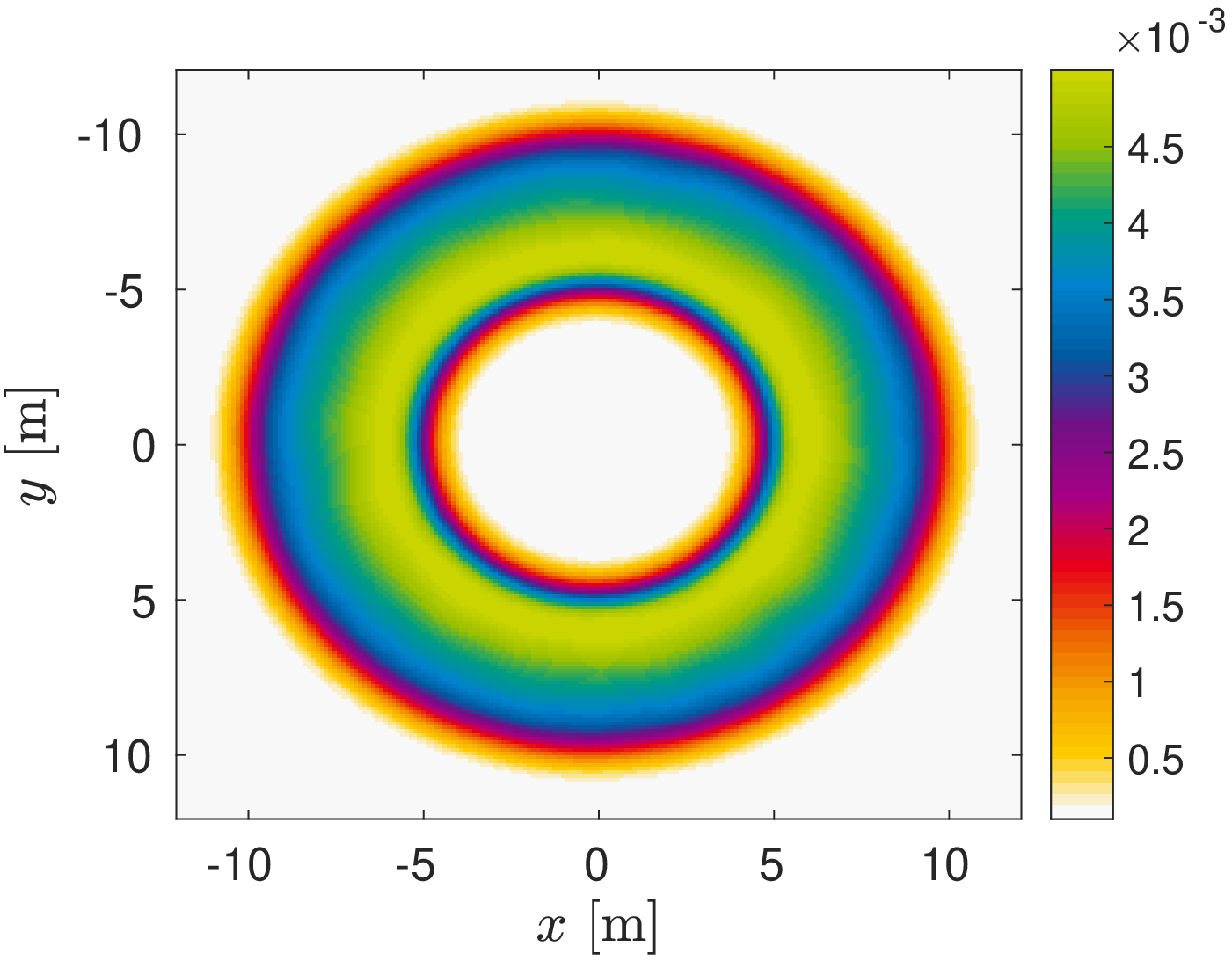}
%  \caption{}
 \label{fig:heuristic sampler_broad}
\end{subfigure}
\caption{\small{A comparison between the proposed sampler (middle) and the heuristic sampler (right) versus the groundtruth (left) for sampling \\$\mathbf{x}_i$ from $Z^{-1}f(r_{ij} \big| \mathbf{x}_i, \mathbf{x}_j^{l'} )$ for the measurement model $r_{ij} = d_{ij} + v$ with $d_{ij} = 7.5$ and $v \sim \mathcal{U}\left[ -2.5, 2.5\right]$.}}
 \label{fig:example of sampling according to IS and VT}
\end{figure*}
The computational complexities for updating messages and beliefs are summarized in \cref{tab:complexity of message update for alpha,tab:complexity of belief update}, respectively. For updating $m_{ij}(\alpha)$ and $m_{ij}(\mathbf{x}_i)$, the BP is computationally substantially more intensive than the SPAWN, see \cref{tab:complexity of message update for alpha}. Regarding updating $B(\mathbf{x}_i)$ using the importance sampler, calculating the importance weights is computationally the most intensive step, requiring operations of order $\mathcal{O}(|\Gamma_i|\cdot L^2 )$. Thanks to the introduction of the auxiliary variable $\boldsymbol{\psi}$ in the proposed AIS, the quadratic complexity order is reduced to the linear order $\mathcal{O}(|\Gamma_i|\cdot L )$, see Table~\ref{tab:complexity of belief update}. 
%%%%%%%%%%%%%%%%%%%%%%%%%%%%%%%%%%%%%%%%%%%%%%%%%%%%%%%%%%%%%%%%%%%%%%%%%%%%%%%%%%%%%%%%%%%%%%%%%%%%%%%%%%%%%%%%%%%%%%%%%%%%%%%%%%%%%%%%%%%%%%%%%%%%%%%%%%
\subsection{Sampling From a Normalized Likelihood Function}
\label{sec:sampling from a normalized likelihood function}

%%%%%%%%%%%%%%%%%%%%%%%%%%%%%%%%%%%%%%%%%%%%%%%%%%%%%%%%%%%%%%%%%%%%%%%%%%%%%%%%%%%%%%%%%%%%%%%%%%%%%%%%%%%%%%%%%%%%%%%%%%%%%%%%%%%%%%%%%%%%%%%%%%%%%%%%
In this subsection, we return to the problem that we have addressed in developing the AIS in \cref{sec:AIS}. For the general measurement model in \cref{eq:measurement model simplified}, our purpose is to sample from the normalized likelihood function $Z^{-1}f(r_{ij} \big| \mathbf{x}_i, \mathbf{x}_j^{l'} )$. The sampling strategy proposed by us is essentially an importance sampler combined with random variable transformation. With the help of random variable transformation, the position samples are generated according to \cref{eq:position samples in existing work}, and the associated proposal distribution $q(\mathbf{x}_i | \mathbf{x}_j^{l'}, r_{ij})$ is derived. In the context of this sampling problem, an importance weight, denoted by $w(\mathbf{x}_i^l)$, is assigned to the sample $\mathbf{x}_i^l$, as given by
 \begin{align}
  w(\mathbf{x}_i^l) \propto \frac{f(r_{ij} \big| \mathbf{x}_i^l,  \mathbf{x}_j^{l'} ) }{ q( \mathbf{x}_i^l \big| \mathbf{x}_j^{l'}, r_{ij}) } = \frac{ f(r_{ij} \big| d_{ij}^l ) }{ q_d( d_{ij}^l \big| r_{ij} )} \cdot d_{ij}^l ,
  \label{eq:weight in IS with random variable transformation}
\end{align}
% \end{subequations}
where $f(r_{ij} \big| d_{ij}^l)$ is $f(r_{ij} \big| \mathbf{x}_i^l,  \mathbf{x}_j^{l'} )$ with $\lVert \mathbf{x}_i^l - \mathbf{x}_j^{l'}\rVert$ replaced by $d_{ij}^l$.
This sampling strategy is related to a heuristic sampling strategy in \cite{Ihler2005}. A straightforward extension of this heuristic sampler leads to the same sample-generating mechanism, i.e., \cref{eq:position samples in existing work}. But different from our sampler, these samples are deemed as following the normalized likelihood function, $Z^{-1}f(r_{ij} \big| \mathbf{x}_i, \mathbf{x}_j^{l'} )$, irrespective of the fact that they actually follow $q(\mathbf{x}_i \big| \mathbf{x}_j^{l'}, r_{ij})$. A question that naturally arises is under which condition are $Z^{-1}f(r_{ij} \big| \mathbf{x}_i, \mathbf{x}_j^{l'} )$ and $q(\mathbf{x}_i \big| \mathbf{x}_j^{l'}, r_{ij})$ proportional. Referring to the relation in \cref{eq:weight in IS with random variable transformation}, it holds only under the condition
\begin{align}
 f(r_{ij} | d_{ij}) \propto q_d(d_{ij} | r_{ij}) / d_{ij}.
\label{eq:condition}
\end{align}
Unfortunately, this condition is not fulfilled in general, and, hence, the heuristic sampler may suffer from performance loss. 

Next, we will compare these two samplers in a concrete example. Consider the measurement model $r_{ij} = d_{ij} + v$ with the true distance $d_{ij} = 7.5$ and the measurement error $v \sim \mathcal{U}\left[ -2.5, 2.5\right]$. 
From the theoretical perspective, the condition in \cref{eq:condition} is not fulfilled here, since we have $f(r_{ij} \left| d_{ij} \right.) = q_{d}(d_{ij} \left| r_{ij} \right.) = f_{v}(r_{ij}-d_{ij})$.
This is also visible in \cref{fig:example of sampling according to IS and VT}, where the kernel density estimate of the proposed sampling strategy and that of the heuristic strategy are depicted, along with the groundtruth $Z^{-1}f(r_{ij} \left| \mathbf{x}_i, \mathbf{x}_j^{l'} \right.)$ in the left plot.
Our sampler, see the middle plot in \cref{fig:example of sampling according to IS and VT},
reflects the groundtruth closely. In contrast, the heuristic sampler, see the right plot in \cref{fig:example of sampling according to IS and VT},
deviates from the groundtruth considerably. Our sampler surpasses the heuristic sampler, in particular, when the likelihood function covers a broad range. On the other hand, when the likelihood function is quite sharp, both samplers can provide quite satisfying approximation results. 
%%%%%%%%%%%%%%%%%%%%%%%%%%%%%%%%%%%%%%%%%%%%%%%%%%%%%%%%%%%%%%%%%%%%%%%%%%%%%%%%%%%%%%%%%%%%%%%%%%%%%%%%%%%%%%%%%%%%%%%%%%%%%%%%%%%%%%%%%%%%
%+++Simulation_Result+++
\section{Simulation Results}
\label{sec:simulation results}
In this section, the performance of the proposed algorithms will be evaluated comprehensively. 
As a comparative algorithm, Tomic's semidefinite programming (SDP) estimator in \cite{Tomic2015} is chosen, since it is shown to outperform the others, including the works in \cite{Vaghefi2013} and \cite{Wang2012}. Here, the SDP estimator is slightly adjusted so that the PLE estimate is constrained in the predefined region, in accordance with $f(\alpha)$. Note that such an adjustment can improve the original SDP estimator, since unreasonable PLE estimates can be avoided. The SDP estimator terminates, either when $N_\text{max} = 100$ iterations are achieved or when $|C(n)-C(n-1)|/|C(n-1)|$ is smaller than $10^{-5}$, where $C(n)$ is the logarithm of the cost function in the $n$-th iteration. The convex optimization problem in the SDP estimator is solved using the CVX Toolbox \cite{cvx} with the SeDuMi solver. 
In the proposed algorithms, the maximal number of iterations is set to $N_\text{max} = 10$, $L = 1000$ particles are used, and $R = 100$ grid points $\{\alpha_d^r\}_{r = 1}^R$ are chosen. 
For a fair comparison with the SDP estimator,  in the proposed algorithms, a point estimate is further inferred from the marginal posterior estimate for each unknown parameter. This is done by finding the highest mode of the analytical form of $B(\mathbf{x}_i)$, which is recovered using kernel density estimation, based on the samples of $B(\mathbf{x}_i)$.
Due to the fact that both the BP-IS and the BP-AIS are computationally very intensive, we will only demonstrate the performance of the SPAWN-IS and that of the SPAWN-AIS. 
%%%%%%%%%%%%%%%%%%%%%%%%%%%%%%%%%%%%%%%%%%%%%%%%%%%%%%%%%%%%%%%%%%%%%%%%%%%%%%%%%%%%%%%%%%%%%%%%%%%%%%%<

\begin{figure}[!htb]
\begin{subfigure}{0.48\linewidth} 
\centering
    \begin{tikzpicture}
      \begin{axis}[
        legend style={
        at={(1,0.5)},
        anchor=east},
        legend cell align=left,
        width= 1.25\textwidth, 
        enlargelimits=false,
      ]
		\pgfplotsset{every tick label/.append style={font=\scriptsize}}
        \addplot+[only marks, mark = square*, mark size = 2 pt] table[x={x}, y={y}] {network_poor_anchor.txt};
        \addlegendentry{\scriptsize{anchor}}
        \addplot+[only marks, mark = *, mark size = 2 pt] table[x={x}, y={y}] {network_poor_agent.txt};
        \addlegendentry{\scriptsize{agent}}

    \end{axis}
  \end{tikzpicture}
%   \caption{}
\label{fig:network I}
\end{subfigure}
\begin{subfigure}{0.5\linewidth} 
\centering
    \begin{tikzpicture}
      \begin{axis}[
%         grid,
        legend style={
        at={(1,0.83)},
        anchor=east},
        legend cell align=left,
        width= 1.2 \linewidth, 
        enlargelimits=false,
      ]
		\pgfplotsset{every tick label/.append style={font=\scriptsize}}
        \addplot+[only marks, mark = square*, mark size = 2 pt] table[x={x}, y={y}] {network_3_anchor.txt};
        \addlegendentry{\scriptsize{anchor}}
        \addplot+[only marks, mark = *, mark size = 2 pt] table[x={x}, y={y}] {network_3_agent.txt};
        \addlegendentry{\scriptsize{agent}}

    \end{axis}
  \end{tikzpicture}
%   \caption{}
\label{fig:network II}
\end{subfigure}
\caption{\small{Network layout: Network I (left) and Network II (right)}}
\label{fig:network}
\end{figure}
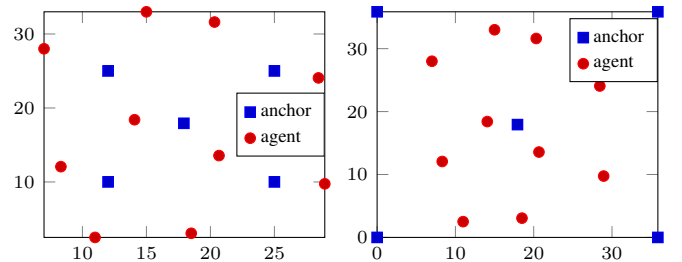

%%%%%%%%%%%%%%%%%%%%%%%%%%%%%%%%%%%%%%%%%%%%%%%%%%%%%%%%%%%%%%%%%%%%%%%%%%%%%%%%%%%%%%%%%%%%%%%%%%%%%%%%%
\begin{figure*}[!htb]
  \begin{subfigure}[b]{0.33\textwidth}
\centering
 \includegraphics[width = 1.0 \textwidth]{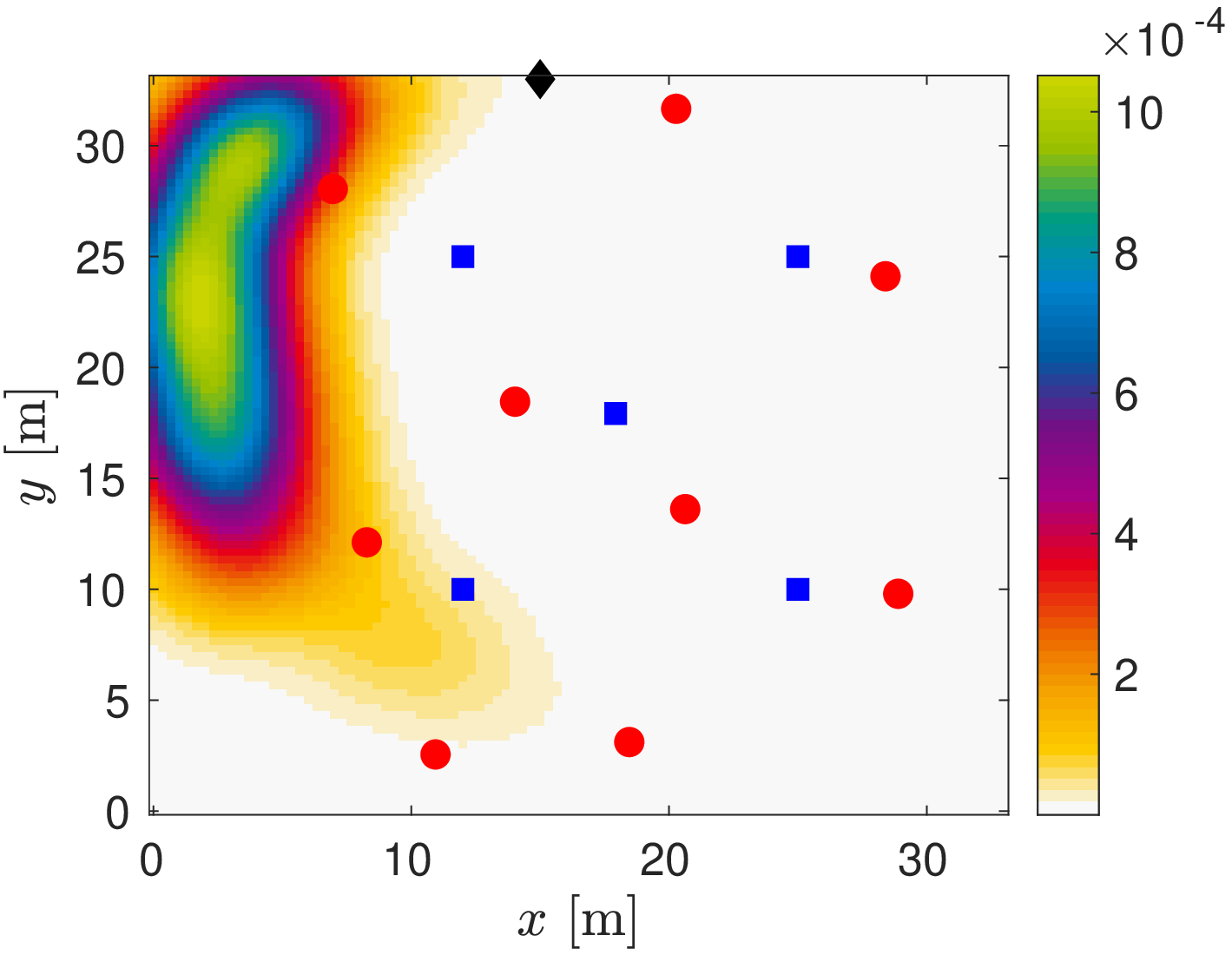}
 \caption{}
 \label{fig:f_x_1}
\end{subfigure}
 \begin{subfigure}[b]{0.33\textwidth}
\centering
 \includegraphics[width = 1.0 \textwidth]{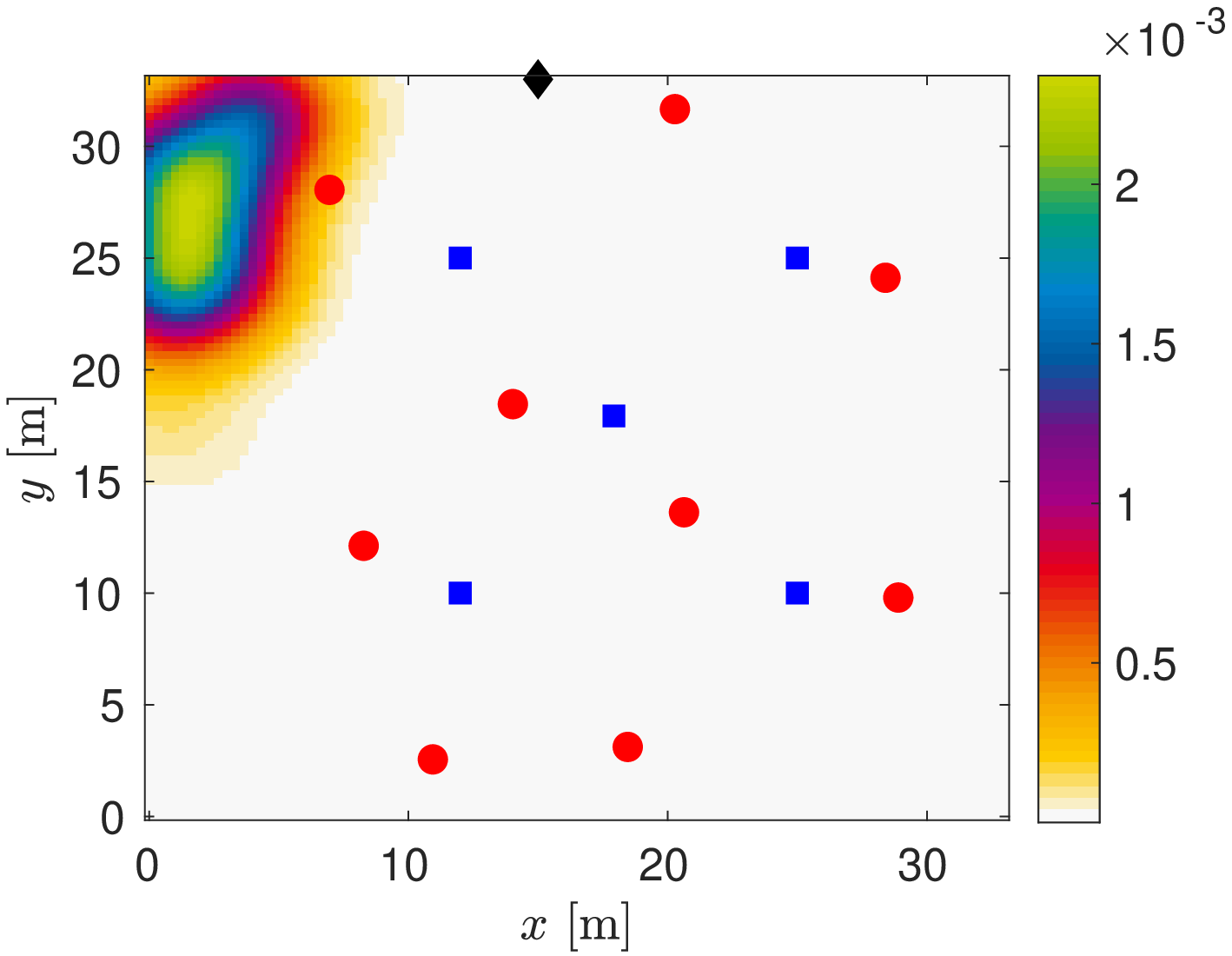}
 \caption{}
 \label{fig:f_x_3}
\end{subfigure}
 \begin{subfigure}[b]{0.33\textwidth}
\centering
 \includegraphics[width = 1.0 \textwidth]{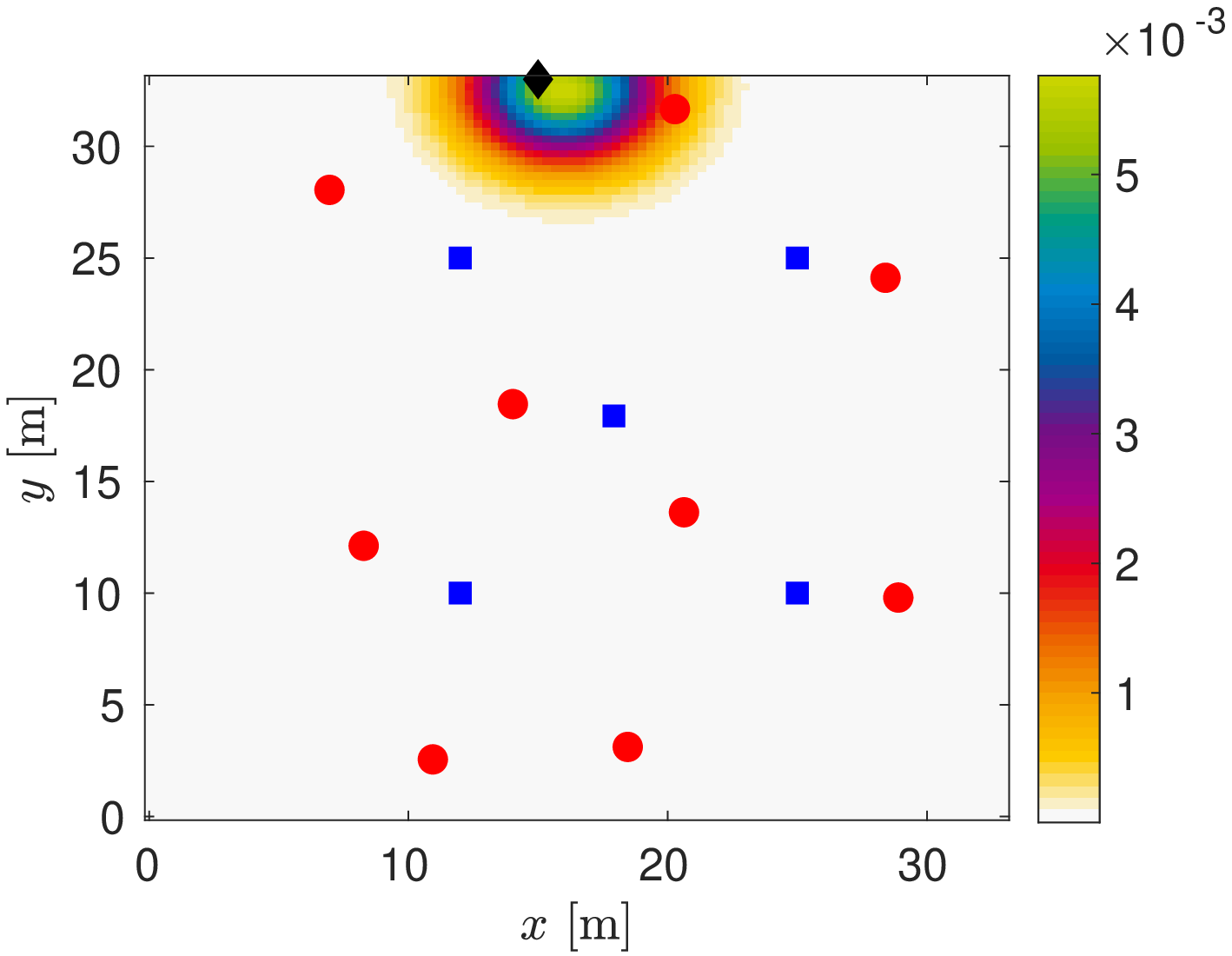}
 \caption{}
 \label{fig:f_x_10}
\end{subfigure}
\caption[]{\small{Example of $B(\mathbf{x}_i)$ in the $1$st (a), $3$rd (b) and $10$-th (c) iteration. The agent of interest locates at \tikz \node[black, fill = black, diamond,inner sep=0.6mm ]{};\,, the anchors locate at \tikz \node[blue, fill = blue, rectangle, inner sep = 1mm]{};\,, and the other agents locate at \tikz\draw[red,fill=red] (0,0) circle (1mm);\,.}}
\label{fig:estimate of marignal posterior}
\end{figure*}

%%%%%%%%%%%%%%%%%%%%%%%%%%%%%%%%%%%%%%%%%%%%%%%%%%%%%%%%%%%%%%%%%%%%%%%%%%%%%%%%%%%%%%%%%%%%%%%%%%%%%%%<

We choose two representative networks with $10$ agents and $5$ anchors: Network I where some of the agents are outside the convex hull of the anchors and Network II where all agents locate within the convex hull, see \cref{fig:network}. The reference power is set to $A_i = -30$ dBm for all $i \in S$ at a reference distance of $d_0 = 1$ meter. For a fair comparison with the SDP estimator, we set the prior distribution of the PLE $\alpha$ as a uniform distribution, $\alpha \sim \mathcal{U}\left[1.5, 6\right]$ and that of each position as a uniform distribution in a square area that is determined by the maximum and the minimum of all nodes' positions. 
All simulation results are based on $100$ Monte Carlo runs. The mean squared error (MSE) of the estimator $\hat{\alpha}$, the bias of $\hat{\alpha}$ and the root mean squared error (RMSE), defined in \cite{Yin2015}, are chosen as performance metrics.

\begin{figure}[!htb]
%  \begin{subfigure}[b]{0.23\textwidth}
\centering
\begin{tikzpicture}
  \begin{axis}[
		grid,
        legend style={
        at={(1,0.7)},
        anchor=east},
        xmin = 3,
        xmax = 4,
        legend cell align=left,
        width = 0.8\linewidth, 
        height = 0.2\textheight,
        enlargelimits=false,
        ylabel={\scriptsize{$B(\alpha)$}}, 
        xlabel={\scriptsize{$\alpha$}},
		xlabel near ticks,
		ylabel near ticks,
      ]
		\pgfplotsset{every tick label/.append style={font=\scriptsize}};
        \addplot+[black, thick, mark = none, dashdotted] table[x={x}, y={y}] {f_alpha_1.txt};
        \addlegendentry{\scriptsize{$1$st}}
		\addplot+[black, thick, mark = none, dashed] table[x={x}, y={y}] {f_alpha_3.txt};
        \addlegendentry{\scriptsize{$3$rd}}
		\addplot+[black, thick, mark = none, solid] table[x={x}, y={y}] {f_alpha_10.txt};
        \addlegendentry{\scriptsize{$10$-th}}
% 		\draw [black, thick, dotted] (3.5, 0) -- (3.5, 10);
 \end{axis}
\end{tikzpicture} 
\caption{\small{Example of $B(\alpha)$ over iterations with the true $\alpha = 3.5$.}}
\label{fig:f_alpha}
% \end{subfigure}
\end{figure}
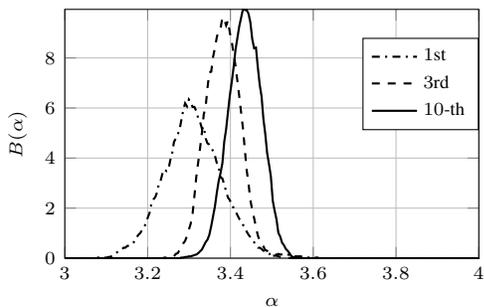
% % % % % % % % % % % % % % % % % % % % % % % % % % % % % % % % % % %
\subsection{Varying Path Loss Exponent}
In this subsection, the purpose is to investigate the performance of the proposed algorithms at different PLE values in different network layouts. We set the standard deviation of the measurement error to $\sigma = 3$ and the communication range to $20$ meter. 
As an illustrative example, we first demonstrate how the beliefs evolve with iterations and depict the kernel density estimates of $B(\mathbf{x}_i)$ and $B(\alpha)$ in \cref{fig:estimate of marignal posterior,fig:f_alpha}, respectively.
It is observed in \cref{fig:estimate of marignal posterior} that over iterations $B(\mathbf{x}_i)$ becomes more concentrated and shifts towards the true position. 
Similarly, over iterations, the uncertainty on $\alpha$ reduces, and $B(\alpha)$ moves towards the true PLE $\alpha = 3.5$, see \cref{fig:f_alpha}. It is noteworthy that the prior distributions adopted are quite coarse, for instance, a uniform distribution $ \mathcal{U}\left[1.5, 6\right]$ is used for the PLE variable. Even so, the proposed algorithms can provide marginal posterior estimates that are relatively sharp and close to the true parameters. 

\begin{figure}[!htb]
\begin{subfigure}[c]{0.49\linewidth}
 \centering
 \tikzstyle{style SDP} = [solid]
 \tikzstyle{style SPAWN} = [solid]
    \begin{tikzpicture} 
      \begin{axis}[
      grid,
        legend style={
        at={(1, 1)},
        anchor=east},
        legend cell align=left,
        width= 1.2\linewidth, 
        enlargelimits=false,
      ]
		\pgfplotsset{every tick label/.append style={font=\scriptsize}}
        \addplot+[only marks, mark = square*, mark color = blue, mark size = 2 pt] table[x={x}, y={y}] {anchor.txt};
%         \addlegendentry{\scriptsize{anchor}}
        \addplot+[only marks, mark = *, mark color = red, mark size = 2.5 pt] table[x={x}, y={y}] {agent.txt};
        \addplot+[only marks, mark = triangle*, mark color = black, mark size = 3 pt] table[x={x}, y={y}] {position_estimate_SPAWN-AIS.txt};
%         \addlegendentry{\scriptsize{SPAWN-AIS}}
        
        %% draw line connnecting agents and estimated positions
        \coordinate (agent 1) at (axis cs: 8.318573, 12.067195); 
		\coordinate (agent 2) at (axis cs: 7.000000, 28.000000); 
		\coordinate (agent 3) at (axis cs: 10.985717, 2.504516); 
		\coordinate (agent 4) at (axis cs: 15.000000, 33.000000); 
		\coordinate (agent 5) at (axis cs: 14.071792, 18.409607); 
		\coordinate (agent 6) at (axis cs: 18.502847, 3.062930); 
		\coordinate (agent 7) at (axis cs: 20.322322, 31.614010); 
		\coordinate (agent 8) at (axis cs: 20.671522, 13.566139); 
		\coordinate (agent 9) at (axis cs: 28.442789, 24.064806); 
		\coordinate (agent 10) at (axis cs: 28.936394, 9.749019); 

\coordinate (agent SPAWN 1) at (axis cs: 7.333333, 11.666667); 
\coordinate (agent SPAWN 2) at (axis cs: 8.000000, 29.666667); 
\coordinate (agent SPAWN 3) at (axis cs: 10.666667, 2.000000); 
\coordinate (agent SPAWN 4) at (axis cs: 16.333333, 32.333333); 
\coordinate (agent SPAWN 5) at (axis cs: 12.666667, 18.000000); 
\coordinate (agent SPAWN 6) at (axis cs: 16.666667, 1.333333); 
\coordinate (agent SPAWN 7) at (axis cs: 20.000000, 31.000000); 
\coordinate (agent SPAWN 8) at (axis cs: 20.000000, 13.333333); 
\coordinate (agent SPAWN 9) at (axis cs: 25.333333, 21.666667); 
\coordinate (agent SPAWN 10) at (axis cs: 27.333333, 7.000000); 
        
        \draw [style SPAWN] (agent 1) -- (agent SPAWN 1);
        \draw [style SPAWN] (agent 2) -- (agent SPAWN 2);
        \draw [style SPAWN] (agent 3) -- (agent SPAWN 3);
        \draw [style SPAWN] (agent 4) -- (agent SPAWN 4);
        \draw [style SPAWN] (agent 5) -- (agent SPAWN 5);
        \draw [style SPAWN] (agent 6) -- (agent SPAWN 6);
        \draw [style SPAWN] (agent 7) -- (agent SPAWN 7);
        \draw [style SPAWN] (agent 8) -- (agent SPAWN 8);
        \draw [style SPAWN] (agent 9) -- (agent SPAWN 9);
        \draw [style SPAWN] (agent 10) -- (agent SPAWN 10);
    \end{axis}
  \end{tikzpicture}
\end{subfigure}
\begin{subfigure}[c]{0.49\linewidth}
 \centering
 \tikzstyle{style SDP} = [solid]
 \tikzstyle{style SPAWN} = [solid]
    \begin{tikzpicture} 
      \begin{axis}[
      grid,
        legend style={
        at={(1, 1)},
        anchor=east},
        legend cell align=left,
        width= 1.2\linewidth, 
        enlargelimits=false,
      ]
		\pgfplotsset{every tick label/.append style={font=\scriptsize}}
        \addplot+[only marks, mark = square*, mark color = blue, mark size = 2 pt] table[x={x}, y={y}] {anchor.txt};
%         \addlegendentry{\scriptsize{anchor}}
        \addplot+[only marks, mark = *, mark color = red, mark size = 2.5 pt] table[x={x}, y={y}] {agent.txt};
%         \addlegendentry{\scriptsize{agent}}
        \addplot+[only marks, mark = triangle*, mark color = black, mark size = 3 pt] table[x={x}, y={y}] {position_estimate_SDP.txt};
%         \addlegendentry{\scriptsize{SDP}}
        
        %% draw line connnecting agents and estimated positions
        \coordinate (agent 1) at (axis cs: 8.318573, 12.067195); 
		\coordinate (agent 2) at (axis cs: 7.000000, 28.000000); 
		\coordinate (agent 3) at (axis cs: 10.985717, 2.504516); 
		\coordinate (agent 4) at (axis cs: 15.000000, 33.000000); 
		\coordinate (agent 5) at (axis cs: 14.071792, 18.409607); 
		\coordinate (agent 6) at (axis cs: 18.502847, 3.062930); 
		\coordinate (agent 7) at (axis cs: 20.322322, 31.614010); 
		\coordinate (agent 8) at (axis cs: 20.671522, 13.566139); 
		\coordinate (agent 9) at (axis cs: 28.442789, 24.064806); 
		\coordinate (agent 10) at (axis cs: 28.936394, 9.749019); 
 
 \coordinate (agent SDP 1) at (axis cs: 12.283844, 12.929672); 
\coordinate (agent SDP 2) at (axis cs: 11.779280, 21.455349); 
\coordinate (agent SDP 3) at (axis cs: 14.185269, 8.171323); 
\coordinate (agent SDP 4) at (axis cs: 17.558979, 24.739123); 
\coordinate (agent SDP 5) at (axis cs: 16.538273, 17.497759); 
\coordinate (agent SDP 6) at (axis cs: 18.472798, 8.774995); 
\coordinate (agent SDP 7) at (axis cs: 19.007519, 25.438313); 
\coordinate (agent SDP 8) at (axis cs: 20.264037, 14.480039); 
\coordinate (agent SDP 9) at (axis cs: 25.342540, 22.105196); 
\coordinate (agent SDP 10) at (axis cs: 24.990705, 13.371512); 

        \draw [style SDP] (agent 1) -- (agent SDP 1);
        \draw [style SDP] (agent 2) -- (agent SDP 2);
        \draw [style SDP] (agent 3) -- (agent SDP 3);
        \draw [style SDP] (agent 4) -- (agent SDP 4);
        \draw [style SDP] (agent 5) -- (agent SDP 5);
        \draw [style SDP] (agent 6) -- (agent SDP 6);
        \draw [style SDP] (agent 7) -- (agent SDP 7);
        \draw [style SDP] (agent 8) -- (agent SDP 8);
        \draw [style SDP] (agent 9) -- (agent SDP 9);
        \draw [style SDP] (agent 10) -- (agent SDP 10);
        
    \end{axis}
  \end{tikzpicture}
\end{subfigure}
\caption[]{\small{Example of position estimates obtained by the SPAWN-AIS (left) and by the SDP estimator (right) with anchors \tikz \node[blue, fill = blue, rectangle, inner sep = 1mm]{};, agents \tikz\draw[red,fill=red] (0,0) circle (1mm); estimated agents \tikz\node[mark size=4pt,color=brown] {\pgfuseplotmark{triangle*}}; .}\vspace{-5mm}}
\label{fig:position_estimate}
\end{figure}
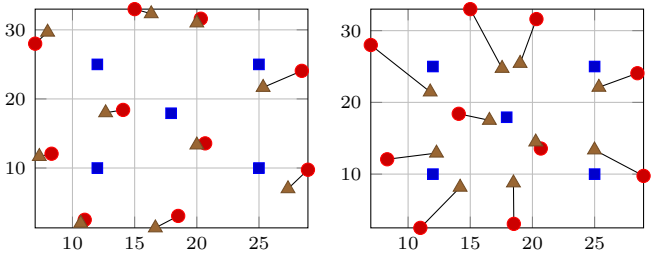

The overall performance of different algorithms is evaluated in terms of the MSE of $\hat{\alpha}$, the bias of $\hat{\alpha}$ and the RMSE, and the results are depicted in \cref{fig:networkI_overAlpha,fig:networkII_overAlpha} for Networks I and II, respectively. For Network I, it is remarkable that, as compared to the SPAWN-IS, the SPAWN-AIS provides comparable performance for both the PLE $\alpha$ and the position $\mathbf{x}_i, i\in S_u$, though its computational complexity is significantly lower. As compared to the SDP estimator, both the localization accuracy and the estimation accuracy of $\alpha$ are improved largely in the proposed algorithms. For a better visualization, we depict the representative position estimates obtained from the SPAWN-AIS and that from the SDP estimator in \cref{fig:position_estimate}. From this figure, it is clear to see that in the SDP estimator the localization accuracy is quite low for the agents outside the convex hull of the anchors, while the SPAWN-AIS does not suffer from this problem. We notice that this type of network topology is rarely examined in the existing literature, although its existence is very probable in practical sensor networks. 
For Network II, again, the MSE curve of $\hat{\alpha}$ in the proposed algorithms is under that of the SDP estimator, see \cref{fig:networkII_overAlpha}, revealing that the proposed algorithms have quite stable estimation performance for the PLE $\alpha$. 
However, for this network, the localization accuracy of the proposed algorithms is comparable to or slightly lower than that of the SDP estimator. This localization performance degradation in the proposed algorithms results from a biased estimation of $\alpha$, which can be seen in the plot on the bottom left in \cref{fig:networkII_overAlpha}. The possible reason for this biased estimation is that there could be certain performance loss when we infer the unknown parameter from its marginal posterior, instead of jointly inferring all unknown parameters from the joint posterior.
Nevertheless, this problem will be alleviated either when the communication range increases or when the measurement noise decreases, as will be demonstrated in the following simulations. 
 
%%%%%%%%%%%%%%%%%%%%%%%%%%%%%%%%%%%%%%%%%%%%%%%%%%%%%%%%%%%%%%%%%%%%%%%%%%%%%%%%%%%%%%%%%%%%%%%%%%%%%%%%%%%%%%%%%
%% network I: RMSE, bias(alpha), MSE(alpha) over alpha
\begin{figure*}[!htb]
	\begin{tabular}{cc}
	 \begin{minipage}{0.48\textwidth} 
	 
	\begin{tabular}[c]{{@{}r@{}}}
	    	\begin{tikzpicture}
		\begin{axis}[
		grid,
        width = 1\textwidth, 
        height = 0.115\textheight,
        enlargelimits=false,
        ylabel={\scriptsize{MSE of $\hat{\alpha}$}}, 
        xlabel={\scriptsize{$\alpha$}},
        xtick={1.5,2,3,4,5,6},
		xticklabels={$1.5$, $2$, $3$, $4$, $5$, $6$}
      ]
		\pgfplotsset{every tick label/.append style={font=\scriptsize}};
		
        \addplot+[black, mark = x, mark size = 3 pt] table[x={x}, y={y}] {network_poor_overAlpha_MSEAlpha_SDP3.txt};
		\addplot+[red, mark = oplus*, mark options=red, mark size = 3 pt] table[x={x}, y={y}] {network_poor_overAlpha_MSEAlpha_SPAWN_prior.txt};
% %        
        \addplot+[blue, mark = triangle*, mark options=blue, mark size = 3 pt] table[x={x}, y={y}] {network_poor_overAlpha_MSEAlpha_SPAWN_auxiliary.txt};
		\end{axis}
		\end{tikzpicture}
		\\
		\begin{tikzpicture}
		\begin{axis}[
	grid,
        width = 1\textwidth, 
        height = 0.115\textheight,
        enlargelimits=false,
        ylabel={\scriptsize{Bias of $\hat{\alpha}$}}, 
        xlabel={\scriptsize{$\alpha$}},
        xtick={1.5,2,3,4,5,6},
		xticklabels={$1.5$, $2$, $3$, $4$, $5$, $6$}
      ]
		\pgfplotsset{every tick label/.append style={font=\scriptsize}};
		
        \addplot+[black, mark = x, mark size = 3 pt] table[x={x}, y={y}] {network_poor_overAlpha_BiasAlpha_SDP3.txt};
%         \addlegendentry{\scriptsize{SDP}}
%         
		\addplot+[red, mark = oplus*, mark options=red, mark size = 3 pt] table[x={x}, y={y}] {network_poor_overAlpha_BiasAlpha_SPAWN_prior.txt};
% %        
        \addplot+[blue, mark = triangle*, mark options=blue, mark size = 3 pt] table[x={x}, y={y}] {network_poor_overAlpha_BiasAlpha_SPAWN_auxiliary.txt};
		\addplot[mark=none, black, dotted, thick] coordinates {(1.5,0) (6,0)};
		\end{axis}
		\end{tikzpicture}
	\end{tabular}
	\end{minipage}
&
    \begin{minipage}{0.5\textwidth} 
		\begin{tikzpicture}
		\begin{axis}[
	grid,
        legend style={
        at={(0.8,0.4)},
        anchor=east},
        legend cell align=left,
        width = 1\textwidth, 
        height = 0.205\textheight,
        enlargelimits=false,
        ylabel={\scriptsize{RMSE [m]}}, 
        xlabel={\scriptsize{$\alpha$}},
        xtick={1.5,2,3,4,5,6},
		xticklabels={$1.5$, $2$, $3$, $4$, $5$, $6$}
      ]
		\pgfplotsset{every tick label/.append style={font=\scriptsize}};
		
        \addplot+[black, mark = x, mark size = 3 pt] table[x={x}, y={y}] {network_poor_overAlpha_RMSE_SDP3.txt};
        \addlegendentry{\scriptsize{SDP}}
		\addplot+[red, mark = oplus*, mark options=red, mark size = 3 pt] table[x={x}, y={y}] {network_poor_overAlpha_RMSE_SPAWN_prior.txt};
        \addlegendentry{\scriptsize{SPAWN-IS}}
%         
% %        
        \addplot+[blue, mark = triangle*, mark options=blue, mark size = 3 pt] table[x={x}, y={y}] {network_poor_overAlpha_RMSE_SPAWN_auxiliary.txt};
        \addlegendentry{\scriptsize{SPAWN-AIS}}
% %         
		\end{axis}
		\end{tikzpicture}
    \end{minipage} 
    \end{tabular}
\caption{\small{Network I: The MSE of $\hat{\alpha}$ (top left), the bias of $\hat{\alpha}$ (bottom left) and the RMSE (right) versus the true PLE $\alpha$. Here, the standard deviation of the measurement error is $\sigma = 3$, and the communication range is $20$ meter. }}
\label{fig:networkI_overAlpha}
\end{figure*}
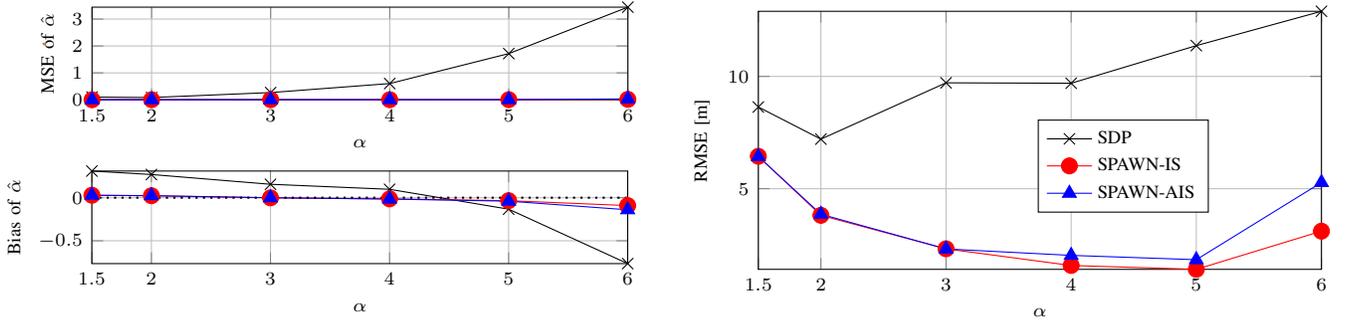
% %%%%%%%%%%%%%%%%%%%%%%%%%%%%%%%%%%%%%%%%%%%%%%%%%%%%%%%%%%%%%%%%%%%%%%%%%%%%%%%%%%%%%%%%%%%%%%%%%%%%%%%%%%%%%%%%%%%%%%% 
%% network II: RMSE, bias(alpha) and MSE(alpha) over alpha
\begin{figure*}[!htb]
	\begin{tabular}{cc}
	 \begin{minipage}{0.48\textwidth} 	 
	\begin{tabular}[c]{{@{}r@{}}}
    	\begin{tikzpicture}
		\begin{axis}[
		grid,
        width = 1\textwidth, 
        height = 0.115\textheight,
        enlargelimits=false,
        ylabel={\scriptsize{MSE of $\alpha$}}, 
        xlabel={\scriptsize{$\alpha$}},
        xtick={1.5,2,3,4,5,6},
		xticklabels={$1.5$, $2$, $3$, $4$, $5$, $6$}
      ]
		\pgfplotsset{every tick label/.append style={font=\scriptsize}};
		
        \addplot+[black, mark = x, mark size = 3 pt] table[x={x}, y={y}] {network_3_overAlpha_MSEAlpha_SDP3.txt};
%         \addlegendentry{\scriptsize{SDP}}
%         
		\addplot+[red, mark = oplus*, mark options=red, mark size = 3 pt] table[x={x}, y={y}] {network_3_overAlpha_MSEAlpha_SPAWN_prior.txt};
% %        
        \addplot+[blue, mark = triangle*, mark options=blue, mark size = 3 pt] table[x={x}, y={y}] {network_3_overAlpha_MSEAlpha_SPAWN_auxiliary.txt};
		\end{axis}
		\end{tikzpicture}
		\\
				\begin{tikzpicture}
		\begin{axis}[
		grid,
%         legend style={
%         at={(0.5,0.3)},
%         anchor=east},
%         legend cell align=left,
        width = 1\textwidth, 
        height = 0.115\textheight,
        enlargelimits=false,
        ylabel={\scriptsize{Bias of $\hat{\alpha}$}}, 
        xlabel={\scriptsize{$\alpha$}},
        xtick={1.5,2,3,4,5,6},
		xticklabels={$1.5$, $2$, $3$, $4$, $5$, $6$}
      ]
		\pgfplotsset{every tick label/.append style={font=\scriptsize}};
		
        \addplot+[black, mark = x, mark size = 3 pt] table[x={x}, y={y}] {network_3_overAlpha_BiasAlpha_SDP3.txt};
%         \addlegendentry{\scriptsize{SDP}}
%         
		\addplot+[red, mark = oplus*, mark options=red, mark size = 3 pt] table[x={x}, y={y}] {network_3_overAlpha_BiasAlpha_SPAWN_prior.txt};
% %        
        \addplot+[blue, mark = triangle*, mark options=blue, mark size = 3 pt] table[x={x}, y={y}] {network_3_overAlpha_BiasAlpha_SPAWN_auxiliary.txt};
		\addplot[mark=none, black, dotted, thick] coordinates {(1.5,0) (6,0)};

		\end{axis}
		\end{tikzpicture}
	\end{tabular}
	\end{minipage}
&
    \begin{minipage}{0.5\textwidth} 
    
		\begin{tikzpicture}
		\begin{axis}[
		grid,
        legend style={
        at={(0.6,0.7)},
        anchor=east},
        ymin = 5,
        ymax = 22,
        legend cell align=left,
        width = 1\textwidth, 
        height = 0.205\textheight,
        enlargelimits=false,
        ylabel={\scriptsize{RMSE [m]}}, 
        xlabel={\scriptsize{$\alpha$}},
        xtick={1.5,2,3,4,5,6},
		xticklabels={$1.5$, $2$, $3$, $4$, $5$, $6$}
      ]
		\pgfplotsset{every tick label/.append style={font=\scriptsize}};
        \addplot+[black, mark = x, mark size = 3 pt] table[x={x}, y={y}] {network_3_overAlpha_RMSE_SDP3.txt};
        \addlegendentry{\scriptsize{SDP}}
		\addplot+[red, mark = oplus*, mark options=red, mark size = 3 pt] table[x={x}, y={y}] {network_3_overAlpha_RMSE_SPAWN_prior.txt};
        \addlegendentry{\scriptsize{SPAWN-IS}}
% %        
        \addplot+[blue, mark = triangle*, mark options=blue, mark size = 3 pt] table[x={x}, y={y}] {network_3_overAlpha_RMSE_SPAWN_auxiliary.txt};
        \addlegendentry{\scriptsize{SPAWN-AIS}}
		\end{axis}
		\end{tikzpicture}
    \end{minipage}
    \end{tabular}
\caption{\small{Network II: The MSE of $\hat{\alpha}$ (top left), the bias of $\hat{\alpha}$ (bottom left) and the RMSE (right) versus the true PLE $\alpha$. Here, the standard deviation of the measurement error is $\sigma = 3$, and the communication range is $20$ meter.}}
\label{fig:networkII_overAlpha}
\end{figure*}
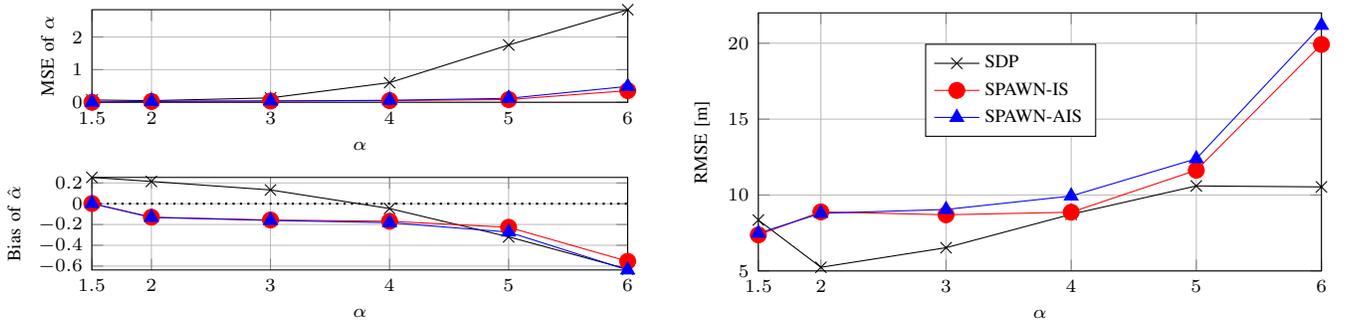
%

%%%%%%%%%%%%%%%%%%%%%%%%%%%%%%%%%%%%%%%%%%%%%%%%%%%%%%%%%%%%%%%%%%%%%%%%%%%%%%%%%%%%%%%%%%%%%%%%%%%%%%%%%%%%%%%%%%%%%%% 
% Example of position estimates
%
 
%%%%%%%%%%%%%%%%%%%%%%%%%%%%%%%%%%%%%%%%%%%%%%%%%%%%%%%%%%%%%%%%%%%%%%%%%%%%%%%%%%%%%%%%%%%%%%%%%%%%%%%%%%%%%%%%%%%%%%% 
% 
\subsection{Varying Communication Range and Standard Deviation}
\label{sec:effect of communcation range}
\begin{figure*}[!htb]
	\begin{tabular}{cc}
    
	 \begin{minipage}{0.48\textwidth} 
	 
	\begin{tabular}[c]{{@{}r@{}}}
	
    	\begin{tikzpicture}
		\begin{axis}[
		grid,
        width = 1\textwidth, 
        height = 0.115\textheight,
        enlargelimits=false,
        ylabel={\scriptsize{MSE of $\hat{\alpha}$}}, 
        xlabel={\scriptsize{$\sigma$}},
        xtick={1,2,3,4,5,6},
		xticklabels={$1$, $2$, $3$, $4$, $5$, $6$},
		ytick = {0, 0.1, 0.2, 0.3},
		yticklabels = {$0$, $0.1$, $0.2$, $0.3$}
      ]
		\pgfplotsset{every tick label/.append style={font=\scriptsize}};
		
        \addplot+[black, mark = x, mark size = 3 pt] table[x={x}, y={y}] {network_3_overSTD_MSEAlpha_SDP3.txt};
		\addplot+[red, mark = oplus*, mark options=red, mark size = 3 pt] table[x={x}, y={y}] {network_3_overSTD_MSEAlpha_SPAWN_prior.txt};
        \addplot+[blue, mark = triangle*, mark options=blue, mark size = 3 pt] table[x={x}, y={y}] {network_3_overSTD_MSEAlpha_SPAWN_auxiliary.txt};
		\end{axis}
		\end{tikzpicture}
	\\
		\begin{tikzpicture}
		\begin{axis}[
		grid, 
        width = 1\textwidth, 
        height = 0.115\textheight,
        enlargelimits=false,
        ylabel={\scriptsize{Bias of $\hat{\alpha}$}}, 
        xlabel={\scriptsize{$\sigma$}},
        xtick={1,2,3,4,5,6},
		xticklabels={$1$, $2$, $3$, $4$, $5$, $6$}
      ]
		\pgfplotsset{every tick label/.append style={font=\scriptsize}};
		
        \addplot+[black, mark = x, mark size = 3 pt] table[x={x}, y={y}] {network_3_overSTD_BiasAlpha_SDP3.txt};
		\addplot+[red, mark = oplus*, mark options=red, mark size = 3 pt] table[x={x}, y={y}] {network_3_overSTD_BiasAlpha_SPAWN_prior.txt};
        \addplot+[blue, mark = triangle*, mark options=blue, mark size = 3 pt] table[x={x}, y={y}] {network_3_overSTD_BiasAlpha_SPAWN_auxiliary.txt};
		\addplot[mark=none, black, dotted, thick] coordinates {(1,0) (6,0)};
		\end{axis}
		\end{tikzpicture}
\end{tabular}
	\end{minipage}
&
    \begin{minipage}{0.5\textwidth} 
		\begin{tikzpicture}
		\begin{axis}[
        grid,
        legend style={
        at={(0.9,0.3)},
        anchor=east},
        legend cell align=left,
        width = 1\textwidth, 
        height = 0.205\textheight,
        enlargelimits=false,
        ylabel={\scriptsize{RMSE [m]}}, 
        xlabel={\scriptsize{$\sigma$}},
        xtick={1,2,3,4,5,6},
		xticklabels={$1$, $2$, $3$, $4$, $5$, $6$}
      ]
		\pgfplotsset{every tick label/.append style={font=\scriptsize}};
		
        \addplot+[black, mark = x, mark size = 3 pt] table[x={x}, y={y}] {network_3_overSTD_RMSE_SDP3.txt};
        \addlegendentry{\scriptsize{SDP}}
		\addplot+[red, mark = oplus*, mark options=red, mark size = 3 pt] table[x={x}, y={y}] {network_3_overSTD_RMSE_SPAWN_prior.txt};
        \addlegendentry{\scriptsize{SPAWN-IS}}
% %        
        \addplot+[blue, mark = triangle*, mark options=blue, mark size = 3 pt] table[x={x}, y={y}] {network_3_overSTD_RMSE_SPAWN_auxiliary.txt};
        \addlegendentry{\scriptsize{SPAWN-AIS}}
		\end{axis}
		\end{tikzpicture}
    \end{minipage} 
    \end{tabular}
\caption{\small{Network II: The MSE of $\hat{\alpha}$ (top left), the bias of $\hat{\alpha}$ (bottom left) and the RMSE (right) versus the standard deviation of the measurement error. Here, the true underlying PLE is $\alpha = 3$, and the communcation range is $25$ meter.}}
\label{fig:networkII_overSTD}
\end{figure*}
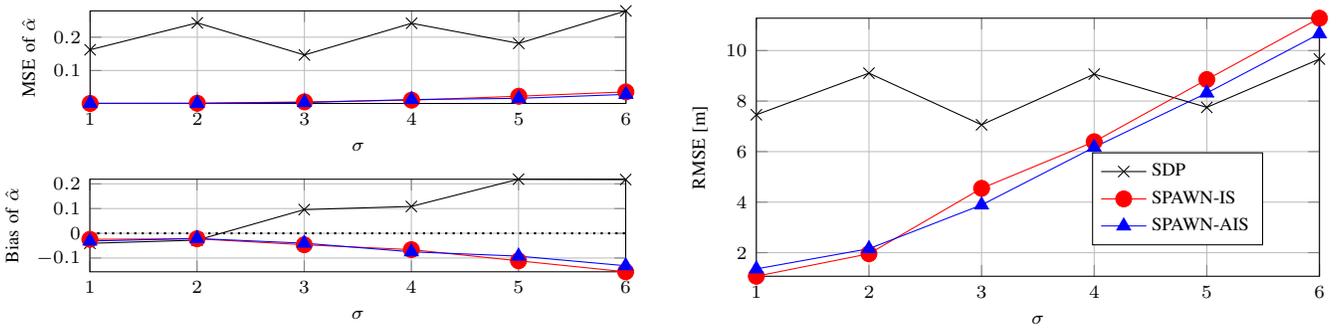

The purpose of this subsection is to assess the performance of the proposed algorithms at varying communication range and varying standard deviation of the measurement error. It has been shown that for Network I the proposed algorithms have quite satisfying performance for both the positions and the PLE. Hence, in the following simulations, we will only focus on Network II.
For the simulation with varying communication range, the true PLE and the standard deviation of the measurement error are set to $\alpha = 3$ and $\sigma = 3$, respectively, and for the other simulation, the true PLE is set to $\alpha = 3$, and the communication range is set to $25$ meter.  
The results are depicted in \cref{fig:networkII_overCommRange,fig:networkII_overSTD} for the cases of varying communication range and varying standard deviation, respectively.

From the figures we can see that in the proposed algorithms all three error-curves drop substantially and eventually attain quite small values, as the communication range increases or the standard deviation of the measurement error decreases. While in contrast, no obvious improvement is seen for the SDP estimator. 
This result is expected and can be explained as follows. 
In the proposed Bayesian algorithms, the marginal posterior of each unknown parameter is inferred. When more information is collected, for instance through increasing communication range (network connectivity) or through decreasing measurement error, the marginal posterior can reflect the unknown parameter more accurately.
On the other hand, the SDP estimator suffers from the performance loss, resulting from the relaxation procedure, and this performance loss may be so dominating that the increase in the information cannot improve the estimation accuracy any more. 
This result highlights that the proposed algorithms can benefit from the increase in the information to a large extent.
Lastly, we stress that although the SPAWN-AIS has a significant reduction on computational cost, it achieves similar estimation performance as the SPAWN-IS.
%%%%%%%%%%%%%%%%%%%%%%%%%%%%%%%%%%%%%%%%%%%%%%%%%%%%%%%%%%%%%%%%%%%%%%%%%%%%%%%%%%%%%%%%%%%%%%%%%%%%%%%%%
 
%% network II: RMSE, bias(alpha), MSE(alpha) over communication range
\begin{figure*}[!htb]
	\begin{tabular}{cc}

	 \begin{minipage}{0.48\textwidth} 
	 
	\begin{tabular}[c]{{@{}r@{}}}
   	\begin{tikzpicture}
		\begin{axis}[
		grid,
        width = 1\textwidth, 
        height = 0.115\textheight,
        enlargelimits=false,
        ylabel={\scriptsize{MSE of $\hat{\alpha}$}}, 
        xlabel={\scriptsize{Communication range [m]}},
        xtick={20, 25, 30, 35, 40},
		xticklabels={$20$, $25$, $30$, $35$, $40$},
		ytick = {0, 0.05, 0.1, 0.15},
		yticklabels = {$0$, $0.05$, $0.1$, $0.15$}
      ]
		\pgfplotsset{every tick label/.append style={font=\scriptsize}};
		
        \addplot+[black, mark = x, mark size = 3 pt] table[x={x}, y={y}] {network_3_overCommRange_MSEAlpha_SDP3.txt};
%         \addlegendentry{\scriptsize{SDP}}
%         
		\addplot+[red, mark = oplus*, mark options=red, mark size = 3 pt] table[x={x}, y={y}] {network_3_overCommRange_MSEAlpha_SPAWN_prior.txt};
% %        
        \addplot+[blue, mark = triangle*, mark options=blue, mark size = 3 pt] table[x={x}, y={y}] {network_3_overCommRange_MSEAlpha_SPAWN_auxiliary.txt};
		\end{axis}
		\end{tikzpicture}
		\\
				\begin{tikzpicture}
		\begin{axis}[
	grid,
        width = 1\textwidth, 
        height = 0.115\textheight,
        enlargelimits=false,
        ylabel={\scriptsize{Bias of $\hat{\alpha}$}}, 
        xlabel={\scriptsize{Communication range [m]}},
        xtick={20, 25, 30, 35, 40},
		xticklabels={$20$, $25$, $30$, $35$, $40$}
      ]
		\pgfplotsset{every tick label/.append style={font=\scriptsize}};
		
        \addplot+[black, mark = x, mark size = 3 pt] table[x={x}, y={y}] {network_3_overCommRange_BiasAlpha_SDP3.txt};
		\addplot+[red, mark = oplus*, mark options=red, mark size = 3 pt] table[x={x}, y={y}] {network_3_overCommRange_BiasAlpha_SPAWN_prior.txt};
% %        
        \addplot+[blue, mark = triangle*, mark options=blue, mark size = 3 pt] table[x={x}, y={y}] {network_3_overCommRange_BiasAlpha_SPAWN_auxiliary.txt};
		\addplot[mark=none, black, dotted, thick] coordinates {(20,0) (40,0)};
		\end{axis}
		\end{tikzpicture}
	\end{tabular}
	\end{minipage}
&
    \begin{minipage}{0.5\textwidth} 
		\begin{tikzpicture}
		\begin{axis}[
		grid,
        legend style={
        at={(0.8,0.3)},
        anchor=east},
        legend cell align=left,
        width = 1\textwidth, 
        height = 0.205\textheight,
        enlargelimits=false,
        ylabel={\scriptsize{RMSE [m]}}, 
        xlabel={\scriptsize{Communication range [m]}},
        xtick={20, 25, 30, 35, 40},
		xticklabels={$20$, $25$, $30$, $35$, $40$}
      ]
		\pgfplotsset{every tick label/.append style={font=\scriptsize}};
		
        \addplot+[black, mark = x, mark size = 3 pt] table[x={x}, y={y}] {network_3_overCommRange_RMSE_SDP3.txt};
        \addlegendentry{\scriptsize{SDP}}
		\addplot+[red, mark = oplus*, mark options=red, mark size = 3 pt] table[x={x}, y={y}] {network_3_overCommRange_RMSE_SPAWN_prior.txt};
        \addlegendentry{\scriptsize{SPAWN-IS}}
% %        
        \addplot+[blue, mark = triangle*, mark options=blue, mark size = 3 pt] table[x={x}, y={y}] {network_3_overCommRange_RMSE_SPAWN_auxiliary.txt};
        \addlegendentry{\scriptsize{SPAWN-AIS}}
% %         
		\end{axis}
		\end{tikzpicture}
    \end{minipage} 
    \end{tabular}
\caption{\small{Network II: The MSE of $\hat{\alpha}$ (top left), the bias of $\hat{\alpha}$ (bottom left) and the RMSE (right) versus the communication range. Here, the standard deviation of the measurement error is $\sigma = 3$, and the true underlying PLE is $\alpha = 3$.}}
\label{fig:networkII_overCommRange}
\end{figure*}
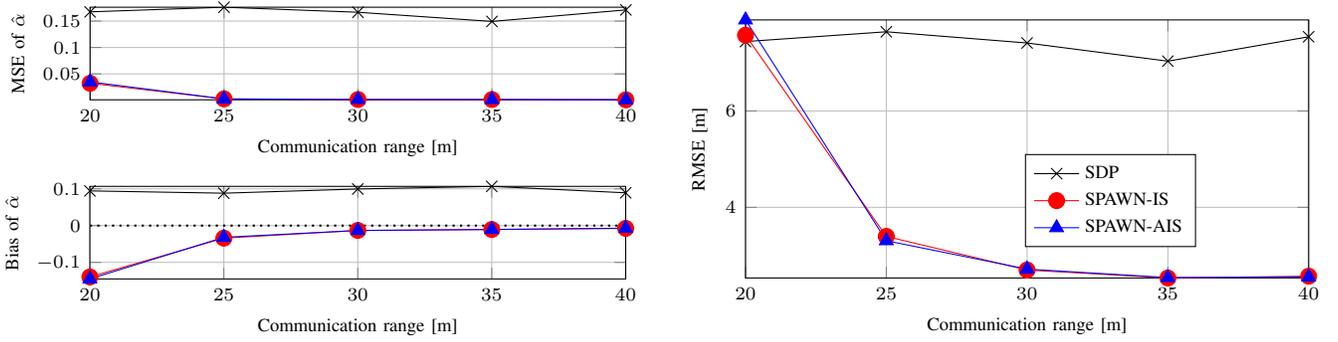
% 
%%%%%%%%%%%%%%%%%%%%%%%%%%%%%%%%%%%%%%%%%%%%%%%%%%%%%%%%%%%%%%%%%%%%%%%%%%%%%%%%%%%%%%%%%%%%%%%%%%%%%%%%%%%%%%%%%%
% 
%%%%%%%%%%%%%%%%%%%%%%%%%%%%%%%%%%%%%%%%%%%%%%%%%%%%%%%%%%%%%%%%%%%%%%%%%%%%%%%%%%%%%%%%%%%%%%%%%%%%%%%%
%+++Conclusion+++
\section{Conclusion}
\label{sec:conclusion}
This paper has proposed a Bayesian framework to address the problem of RSS-based cooperative localization with unknown PLE. To infer the marginal posterior of each unknown parameter, we have developed a series of message passing algorithms. 
The proposed algorithms provide a unified strategy for estimating both the positions and the PLE parameter and, therefore, handle the problem from a more theoretical perspective, as compared to the heuritic alternating strategy in the existing literature. 
The simulation results have demonstrated that: 
$(1)$ As compared to the competitor, the proposed algorithms achieve comparable or better localization performance, depending on the network layout;
$(2)$ The proposed algorithms can benefit from the increase in the information significantly and tend to outperform the existing one in dense networks and low-to-medium noise scenarios;
$(3)$ Concerning the PLE parameter, the proposed algorithms tend to underestimate it, incurring deterioration of localization accuracy. Nevertheless, the proposed algorithms consistently achieve a smaller MSE than the competitor;
$(4)$ Among the proposed algorithms, the SPAWN-AIS achieves comparable performance, but at the lowest computational cost.
Many research challenges need to be overcome in our future work, including reducing the bias in the PLE and extending this work to an inhomogeneous environment.
%%%%%%%%%%%%%%%%%%%%%%%%%%%%%%%%%%%%%%%%%%%%%%%%%%%%%%%%%%%%%%%%%%%%%%%%%%%%%%%%%%
\appendices

%+++Append_02+++
\allowdisplaybreaks
\section{Derivation of \cref{eq:normalization constant}}
\label{sec:integral}% 
%% without detection model
We consider the calculation of the integral in Eq. \eqref{eq:normalization constant}. For notational simplicity, we ignore the superscript $l, n-1$ during this derivation.
The integral is calculated as follows:
\begingroup
\allowdisplaybreaks
\begin{subequations}
 \begin{align*}
  Z_{ij}  	 & =  \int f(r_{ij} | \mathbf{x}_i, \mathbf{x}_j, \alpha) ~\textrm{d}\mathbf{x}_i \\
             & = \int  f_{\mathcal{N}}\left( r_{ij}- A_i + 10\alpha \text{log}_{10}\frac{ \lVert \mathbf{x}_i - \mathbf{x}_j \rVert }{d_0} \right)~\textrm{d} \mathbf{x}_i  \\
             & \overset{ \text{\textcircled{1}} }{=} \int f_{\mathcal{N}}\left( r_{ij}- A_i + 10\alpha \text{log}_{10}\frac{ \lVert \mathbf{x}_{ij} \rVert }{d_0} \right)~\textrm{d} \mathbf{x}_{ij}  \\
             & \overset{ \text{\textcircled{2}} }{=}  \int_{0}^{2\pi} \hspace{-2mm} \int_{0}^{\infty} \hspace{-2mm} f_{\mathcal{N}}\left( r_{ij}- A_i + 10\alpha \text{log}_{10}\frac{ d_{ij} }{d_0} \right)  \cdot d_{ij}  ~\textrm{d} d_{ij} ~\textrm{d} \theta_{ij} \\
             & = 2\pi \frac{\text{log}10}{10\alpha} ~ \mathbb{E}_{d_{ij} \sim \text{log} \mathcal{N}\left(\mu_d, \sigma_d^2 \right) } \left[ d_{ij}^2 \right] \\
%              & = 2\pi \frac{\text{log}10}{10\alpha} \cdot \left( \mu_d^2 +\sigma_d^2 \right) \\
             & = 2\pi \frac{\text{log}10}{10\alpha} \cdot \text{exp}(2\sigma_d^2 + 2\mu_d) 
 \end{align*}
 \end{subequations}
\endgroup
with 
 \begin{align*}
  \mu_d       = \frac{\text{log}10}{10\alpha}(A_i - r_{ij})+\text{log}d_0 , \quad \quad \sigma_d^2  = \sigma^2 \frac{(\text{log}10)^2}{ (10\alpha)^2} .
 \end{align*}
% \end{subequations}
Here, $f_{\mathcal{N}}(\cdot)$ stands for the pdf of the Gaussian distribution $\mathcal{N}(0, \sigma^2)$, \textcircled{1} stands for $\mathbf{x}_{ij} = \mathbf{x}_i - \mathbf{x}_j$, and from \textcircled{1} to \textcircled{2} is achieved by transforming the Cartesian coordinate $\mathbf{x}_{ij}$ to the polar coordinate $\left[d_{ij}, \theta_{ij}\right]^T$.

%%%%%%%%%%%%%%%%%%%%%%%%%%%%%%%%%%%%%%%%%%%%%%%%%%%%%%%%%%%%%%%%%%%%%%%%%%%%%%%%%%%%%%
%+++Append_03+++
 
\section{Derivation of \cref{eq:conditional pdf of distance}}
\label{sec:conditional pdf of distance}
For the measurement model in \cref{eq:measurement model}, the distance sample $d_{ij}^{l}$ generated according to \cref{eq:distance sample,eq:noise sample} fulfills the relation
 \begin{align*}
 \text{log}\frac{d_{ij}^{l}}{d_0}                         & = \underbrace{\frac{\text{log}10}{10\alpha^{l'}} \cdot \left( A_i-r_{ij} \right) + \frac{\text{log}10}{10\alpha^{l'}} \cdot v^l}_{\tilde{v}} \label{eq:lognormal distributed distance}.
\end{align*}
% \end{subequations}
Given $\alpha$, $A_i$ and $r_{ij}$, the variable $\tilde{v}$ is Gaussian distributed, namely, $\tilde{v}\sim \mathcal{N} \left( \tilde{\mu}, \tilde{\sigma}^2 \right)$ with 
\begin{gather*}
 \tilde{\mu}  = \frac{\text{log}10}{10\alpha^{l'}} \cdot \left( A_i - r_{ij} \right), \quad \quad \tilde{\sigma}^2  = \left(\frac{\text{log}10}{10\alpha^{l'}}\right)^2 \cdot \sigma^2 .
\end{gather*}
It follows that $d_{ij}/d_0$ is log-normal distributed, namely, 
 \begin{align*}
 	d_{ij}/d_0 \sim \text{Log}\mathcal{N}\left( \tilde{\mu}, \tilde{\sigma}^2\right).
 \end{align*}
Furthermore, it is given that the pdf of a log-normal-distributed random variable $a \sim \text{Log}\mathcal{N}\left( \mu_a, \sigma_a^2\right)$ is in the form of 
\begin{align}
 f(a)  = \frac{1}{\sqrt{2\pi} a \sigma_a } \text{exp}\left( -\frac{ \left(\text{log}a-\mu_a \right)^2 }{2\sigma_a^2} \right). \label{eq:pdf of lognormal}
\end{align}
Finally, substituting $a$, $\mu_a$ and $\sigma_a$ in \cref{eq:pdf of lognormal} with $d_{ij}/d_0$, $\tilde{\mu}$ and $\tilde{\sigma}^2$, respectively, concludes the derivation.

\section*{Acknowledgment}
The authors would like to thank Mr. Slavisa Tomic for providing the MATLAB codes of the comparative algorithm.

% The authors would like to thank the anonymous reviewers for their helpful and constructive comments that greatly improved the quality of the paper.
\ifCLASSOPTIONcaptionsoff
  \newpage
\fi



%\bibliographystyle{IEEEtran}
% \end{thebibliography}
% 
\bibliographystyle{IEEEtran}
\bibliography{Ref}

% \end{IEEEbiography}


%\newpage



%\vfill


\end{document}